\documentclass[twocolumn]{aastex631}

\usepackage{enumitem}

\usepackage{amsmath}
\usepackage{hyperref}

\begin{document}

\title{Mitigating Random-Phase Sampling Noise in the Cepheid Period-Luminosity Relation: \\A Cross-Filter Consistency Approach}

\shorttitle{Mitigating Random-Phase Sampling Noise in the Cepheid PL Relation}

\author[0009-0006-3280-5622]{Mahdi Abdollahi}
\affiliation{School of Astronomy, Institute for Research in Fundamental Sciences (IPM), P.O. Box 1956836613, Tehran, Iran}

\author[0000-0001-8392-6754]{Atefeh Javadi}
\affiliation{School of Astronomy, Institute for Research in Fundamental Sciences (IPM), P.O. Box 1956836613, Tehran, Iran}

\author[0000-0002-1576-1676]{Barry F. Madore}
\affiliation{The Observatories, Carnegie Institution for Science, 813 Santa Barbara St., Pasadena, CA 91101, USA}
\affiliation{Department of Astronomy $\And$ Astrophysics, University of Chicago, 5640 South Ellis Avenue, Chicago, IL 60637, USA}

\author[0000-0003-3431-9135]{Wendy L. Freedman}
\affiliation{Department of Astronomy $\And$ Astrophysics, University of Chicago, 5640 South Ellis Avenue, Chicago, IL 60637, USA}
\affiliation{Kavli Institute for Cosmological Physics, University of Chicago, 5640 S. Ellis Ave., Chicago, IL 60637, USA}

\author[0009-0002-9957-5818]
{Hamidreza Mahani}
\affiliation{School of Astronomy, Institute for Research in Fundamental Sciences (IPM), P.O. Box 1956836613, Tehran, Iran}

\correspondingauthor{Atefeh Javadi}
\email{atefeh@ipm.ir}

\begin{abstract}
The Period-Luminosity (PL) relation is usually derived using time-averaged magnitudes, which require multiple-epoch observations to determine periods and adequately sample the light curves. Although single-epoch observations are more practical and require significantly less observational effort, they inherently introduce greater photometric scatter, leading to an increased dispersion in the derived Period-Luminosity relations. In this paper, we explore, in detail, a method that transforms single random-phase data to their mean-light values, using information obtained in other bands for the same Cepheid. This approach enables the accurate re-construction of mean-light PL relations for wavelengths observed with space-based facilities, for instance, where the number of epochs per star makes simple averaging or template fitting less than optimal, with the latter requiring very high-precision periods for predictive phasing. While applying this technique across multiple bands, from optical to mid-IR, we focus particularly on widely separated bands covering the mid-IR to the optical. We showcase this method using the J band (as being observed by JWST) as the random-phase component. Our results show that this correction reduces the scatter of the PL relation in the J band by a factor of approximately $0.7\times$, equivalent to increasing the number of random-phase observations by a factor of 10, needed to obtain the same increase in precision as delivered here.

\end{abstract}

\keywords{Galaxy Distances(590); Cepheid distance (217)}

\section{Introduction} \label{sec:intro}


Cepheids are variable stars that are among the best-understood and most well-studied primary distance indicators, used as an important step in the distance ladder \citep{Leavitt_1912_Cep_PLR}. The Large Magellanic Cloud (LMC), serving as a crucial bridge for exploring extragalactic structures and compositions, provides an excellent laboratory for calibrating the PL relation using Cepheid variables \citep{Sandage_Tammann_1968_PLR,Riess_2019_LMC_Cal,Pietrzy_2019_LMC_Cal}.  Comprehensive datasets such as the Surveying the Agents of Galaxy Evolution (SAGE) \citep{SAGE_2006}, the Visible and Infrared Survey Telescope for Astronomy (VISTA) \citep{VISTA_2011} and Optical Gravitational Lensing Experiment (OGLE) \citep{Soszy_2015_OGLE} surveys have played a pivotal role in these studies, having offered extensive spectroscopic and photometric coverage of LMC sources.

Classical Cepheids are located within the instability strip in the Hertzsprung-Russell color-magnitude diagram (CMD). In the optical, Cepheids are characterized by saw-toothed brightness variations, having periods ranging from 3 to over 100 day. These light variations arise from periodic changes in the star’s total surface area and mean surface brightness, which are governed by variations in radius and temperature, respectively. Both parameters are physically linked through the Stefan's law, which relates the star's luminosity to its radius and effective temperature \citep{Madore_freedman_1991,Freedman_2010_multi_LC}. The period, to first order, is largely controlled by the star's mean density.

One of the most significant characteristics of Cepheids is the extensive multiwavelength data available in various astronomical databases. \citet{Madore_freedman_1991} provided early B to K photometry for LMC Cepheids, followed by \citet{Fouqu_2007_MW_data} who compiled similar data for Milky Way Cepheids. The OGLE project later delivered long-term V and I photometry for LMC and SMC Cepheids \citep{Soszy_2015_OGLE}.

\citet{Macri_2015_NIR} extended observations into the near-infrared H-band for LMC Cepheids. Mid-infrared (3.6 and 4.5 $\mu m$) Spitzer photometry was provided by \citet{Scowcroft_2011_Spitzer} for LMC Cepheids and by \cite{Scowcroft_2016_Spitzer_SMC} for SMC Cepheids. Near-infrared YJK photometry for SMC Cepheids was added by \citet{Ripepi_2017_NIR_SMC}.

More recently, \citet{Ripepi_2022_NIR} presented VMC YJKs light curves for LMC Cepheids, and \citet{Bhardwaj_2024_MultiW_CMetall} compiled a comprehensive grizJHK dataset for Milky Way Cepheids. These datasets enable comparisons of observations across multiple filters and facilitate the development of methods to reduce uncertainties by leveraging data from different bands \citep{Madore_2017_EX_DM_Cor,Madore_2024_NovelMethod}.

To mitigate the uncertainties, it is first necessary to examine how various physical parameters influence the luminosity and periodicity of Cepheid variables.  Studies have shown an inverse correlation between wavelength and the intrinsic scatter in the PL relation \citep{Madore_2011_multi,Ripepi_2019_Reclassification_Cep_GaiaDR2,Breuval_2022_Metallicity}. This can be understood through Stefan’s law \citep{Stefan_1879}, where Cepheid luminosity variations are governed by two primary parameters: radius and temperature.

Variations in the radius lead to changes in the stellar surface area, thereby affecting luminosity. However, since the surface area and its changes are independent of wavelength, The influence of radius change should be achromatic, and contribute equally across different bands.

In contrast, the impact of temperature will be wavelength-dependent, as broadly described by Wien’s displacement law \citep{Wien_1896}. At longer wavelengths, the influence of temperature variations upon surface brightness diminishes, leading to reduced Cepheid amplitudes and decreased width in the PL relation in the near and far-infrared \citep{Freedman_2010_multi_LC}. The practical implications are that random sampling of an individual Cepheid's light curve and/or the random sampling of a Cepheid in the period-luminosity relation, will always fall closer to the mean, in the infrared, than those same samplings if they are made in the optical.

As a result, longer wavelengths, particularly in the infrared (IR) regime, have been employed to minimize scatter in the PL relation, thereby reducing errors in its calibration. Moreover, IR wavelengths offer the hope of reduced sensitivity of Cepheid magnitudes and colors to metallicity effects, and they certainly mitigate the impact of interstellar extinction \citep{Freedman_1990_Manifold,Persson_2004_nearIR}.

As a result, longer wavelengths, particularly in the infrared (IR), have been used to reduce the scatter in the PL relation and improve its calibration. Although there is still much debate about how much IR observations can reduce the sensitivity of Cepheid magnitudes and colors to metallicity, it is well established that IR data greatly lessen the effects of interstellar extinction, making the PL relation more reliable \cite{freedman_2012_carnegie,Breuval_2022_Metallicity,Bhardwaj_2024_MultiW_CMetall,Trentin_2024_CMetall}.

As a result, longer wavelengths, particularly in the infrared (IR) regime, have been employed to minimize scatter in the PL relation, thereby reducing errors in its calibration.  Moreover, IR wavelengths suffer significantly less from interstellar extinction compared to optical wavelengths \citep{Freedman_1990_Manifold,Persson_2004_nearIR,freedman_2012_carnegie}.

The importance  of IR observations of Galactic Cepheids for refining the cosmic distance scale became increasingly recognized over time, building upon the pioneering work of \citet{wisniewski_johnson_1968}. The first application of near-infrared observations to study the Cepheid period-luminosity (PL) relation was carried out by \citet{McGonegal_1982_First_NIR_PLR}, who utilized random-phase magnitudes in the H band using the Cerro Tololo Inter-American Observatory (CTIO) InSb photometer. \citet{Welch_1987_Hband_PLR} later provided a comprehensive summary of the early advancements in establishing the PL relation in the near-infrared regime. Building on these early efforts, \citet{Persson_2004_nearIR} presented the first comprehensive set of well-sampled near-infrared ($JHK_s$) light curves for fundamental-mode Cepheids in the Large Magellanic Cloud (LMC), enabling a precise calibration of the PL relation in these bands. \citet{Macri_2015_NIR} expanded the NIR sample by surveying the central regions of the LMC, covering 3.6 million objects, and derived both the Tip of the Red Giant Branch (TRGB) and Cepheid PL relations. More recently, \citet{Ripepi_2022_NIR} presented an extensive NIR catalog based on VISTA observations, constructing a detailed 3D map of the LMC and refining the PL relations with high precision.

However, ground-based observations of mid- and far-infrared wavelengths face significant challenges due to the Earth's atmosphere, which absorbs much of that radiation and introduces foreground thermal noise. To overcome these limitations, space-based telescopes such as \textit{Hubble} Space Telescope's (\textbf{HST}), the \textit{Spitzer} Space Telescope and the \textit{James Webb} Space Telescope (\textit{JWST}) are indispensable \citep{Kimble_2008_NIR_HST,Werner_2004_Spitzer,Gardner_2023_JWST}. Operating above the atmosphere, these telescopes provide uninterrupted access to the IR regime, enabling precise measurements of Cepheid variables.

\cite{Freedman_2008_First_Epoch} initially established PL relations in the mid-infrared regime using first-epoch observations from the SAGE survey, carried out with the \textit{Spitzer} Space Telescope. This work included a sample of 92 stars selected from the \cite{Persson_2004_nearIR} JHK sample of least-crowded LMC Cepheids. Later, \cite{Madore_2009_Second_Epoch} incorporated a second epoch of observations into the dataset, enabling the computation of mean magnitudes across the two epochs for each star and facilitating further refinement of the PL relation.  Their analysis demonstrated that approximately half of the uncertainty in these mid-IR PL relations was attributable to random-phase sampling. Notably, compared to constructing the PL relation using {\bf time-averaged optical} observations, the scatter in the PL relation derived from IR bands, even when affected by random-phase errors, was significantly reduced (see \citet{McGonegal_1982_First_NIR_PLR} for a very early demonstration of this fact.)

The next challenge in achieving more accurate distance measurements lies in mitigating the scatter introduced by random-phase sampling, a type of deterministic, but reducible, statistical error arising from the inherent variability of the Cepheids themselves. These errors can obviously be reduced by simply increasing the number of observations from single to multiple epochs  (e.g., \cite{Madore_2009_Second_Epoch} if additional telescope time can be acquired.)

For ground-based telescopes, repeated observations of a field are feasible and have been successfully implemented in large-scale surveys. For instance, the OGLE (Optical Gravitational Lensing Experiment) \citep{Udalski_1992_OGLE} and MACHO (Massive Compact Halo Object) Projects \citep{Alcock_1997_MACHO} have conducted extensive multi-epoch observations of the Large and Small Magellanic Clouds.

However, for space-based telescopes such as the \textit{HST}, the \textit{JWST} and the \textit{Nancy Grace Roman} Space Telescope (\textit{RST}) \citep{RST_2015}, limited observing time often makes it impractical to repeatedly observe the same field \citep{Madore_2024_NovelMethod}. Nevertheless, certain observational constraints are inherent to the allocation of telescope time. For instance, it is often not feasible to schedule multiple observations of the same field due to the large number of competing targets. Despite these limitations, advanced instrumentation and analysis techniques help mitigate their impact. In particular, for Cepheids and their PL relation, it is possible to correct for random-phase observations and derive accurate mean magnitudes from single-epoch data. This approach ensures that the scientific return of each observation is maximized, while preserving the essential role of Cepheids in calibrating the cosmic distance ladder.
\begin{figure*}
    \centering
    \begin{minipage}{0.8\textwidth}
        \includegraphics[width=0.49\linewidth]{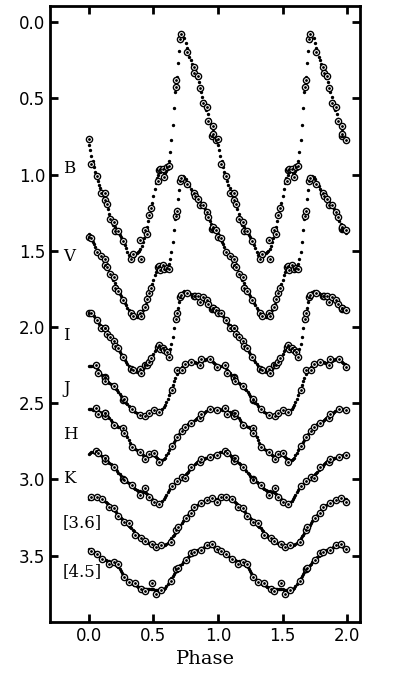}
        \includegraphics[width=0.49\linewidth]{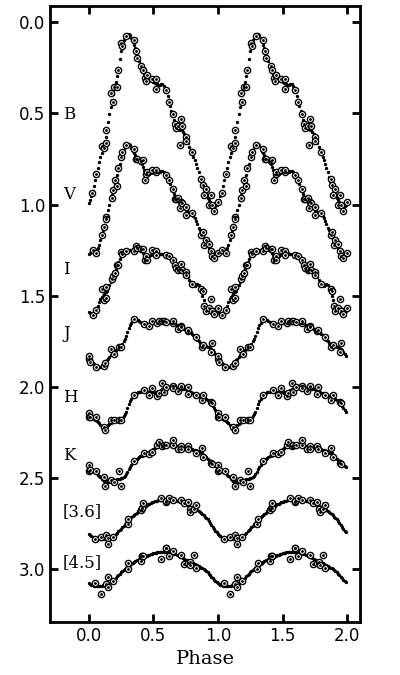}
    \end{minipage}
    \caption{Multiwavelength light curves of the Cepheid variable stars HV2324 (left panel) and HV2854 (right panel) in the LMC, with periods of $14.4657$ and $8.6348$ days, respectively, are presented. The data were obtained using the Three-hundred MilliMeter Telescope (TMMT; Monson et al. 2025, in preparation), the 1-m Swope and 2.5-m duPont telescopes \citep{Persson_2004_nearIR}, and the Spitzer Space Telescope \citep{Scowcroft_2011_Spitzer}, covering optical, near-IR, and mid-IR wavelengths, respectively. The light curves are represented by circled dots and have been phase-folded using the \texttt{Lightkurve} algorithm \citep{LightKurve}. The solid black dots correspond to the smoothed, uniformly distributed phase-magnitude data generated using the GLOESS method, as described in Appendix~\ref{sec:GLOESS}.}
    \label{fig:Single_All_filters}
\end{figure*}
\textit{JWST}, with its exceptional infrared sensitivity and high spatial resolution, significantly reduces the effects of stellar crowding, which is a major source of uncertainty in the PL relation. However, some studies based on high-resolution optical data suggest that the impact of crowding may be less significant than previously assumed \citep{Gibson_2000, Anderson_2018}.  By resolving individual Cepheids in distant galaxies and minimizing contamination from neighboring stars, \textit{JWST} enables more precise measurements of their luminosities, refining the calibration of the distance scale \citep{Freedman_2024_IAUS376,Riess_2024_JWST}. Similarly, the RST, equipped with a wide field camera and capable of high-precision photometry, is designed for large-scale surveys, enabling the efficient detection of Cepheids across numerous galaxies\citep{spergel_2015_RST}.

To address the challenge of deriving Cepheid mean magnitudes from single-epoch observations, a variety of methods have been proposed.

One of the earliest approaches relied on template light curves. \citet{Welch84} suggested scaling the amplitudes of Galactic Cepheid light curves with similar periods and using them as templates to estimate mean JHK magnitudes. \citet{Freedman88} applied a related method in the optical, deriving V- and I-band light curves for Cepheids in IC 1613 by scaling amplitudes and shifting phases of B-band light curves. This approach relied on empirically determined amplitude ratios and phase lags from Galactic Cepheids and assumed that light-curve shapes were the same across filters. \citet{Stetson96} later introduced Fourier-based template light curves constructed from large samples of Galactic and Magellanic Cloud Cepheids, a technique later applied to Hubble Space Telescope observations of sparsely sampled Cepheids (e.g., \citet{Gibson2000}).

Other methods aimed to avoid strict assumptions about light-curve shapes. \citet{Labhardt97} derived empirical correction curves to transfer light curves between filters (V to B, R, I), without relying on fixed template forms. \citet{Ngeow03} reconstructed I-band light curves from V-band data by exploiting statistical correlations between Fourier parameters, a method that requires precise Fourier coefficients up to fourth order and therefore depends on high-quality, well-sampled V-band light curves. \citet{Nikolaev04} introduced a different strategy for the NIR: using more than 2000 LMC Cepheids, they derived correction functions by comparing observed magnitudes with those predicted from PL relations and fitting the residuals as a function of V-band phase. While effective, this method does not account for amplitude variations among Cepheids and typically results in uncertainties of about 0.05 mag, similar to other techniques.

In the near-infrared, additional strategies have been developed to estimate mean magnitudes from sparse data. \citet{Soszynski05} developed a procedure in which light-curve templates are fitted with multiple free parameters to estimate mean magnitudes, with phase lags between filters used to transform optical templates into their NIR counterparts. More recently, \citet{Riess23} applied a similar template-based approach, again relying on phase lags to recover mean magnitudes from limited-epoch data.

A complementary approach has been introduced by \citet{Madore_2024_NovelMethod}, who developed a systematic framework for recovering mean Cepheid magnitudes from single-epoch data by combining information across multiple bands. Their method constructs normalized residuals of the PL relations in well-sampled filters, ranking Cepheids according to their relative positions within the instability strip. These rankings are then averaged across filters to form a cross-band “template ranking,” which can be scaled and applied to single-epoch data in a target band to predict the mean magnitudes. This approach effectively reduces phase-induced scatter by leveraging multi-wavelength correlations.

The method employed in this work addresses random-phase sampling noise in the PL relation by making use of direct correlations between residuals in different bands. Two types of correlations are considered: (i) the residuals of the PL relation in one band compared with those in another band, and (ii) the residuals of the PL relation in one band compared with the magnitude–magnitude residuals between two bands. Together, these correlations allow the mean magnitude of a Cepheid in a sparsely sampled band to be estimated if its mean magnitude is known in a companion band. Hence, the method introduced here is novel and can be applied even when only one other band has a well-determined mean magnitude. As a result, it provides a straightforward and robust way to recover mean magnitudes from single-epoch data, while reducing systematic errors caused by incomplete phase coverage.

The structure of this paper is as follows. Section \ref{sec:data} presents the data used in this work, Section \ref{sec:methodology}  details the methodology, and Section \ref{sec:results} reports the results. The discussion and conclusions are given in Section \ref{sec:conlusion}.
\begin{figure}
    \centering
    \includegraphics[width=\columnwidth]{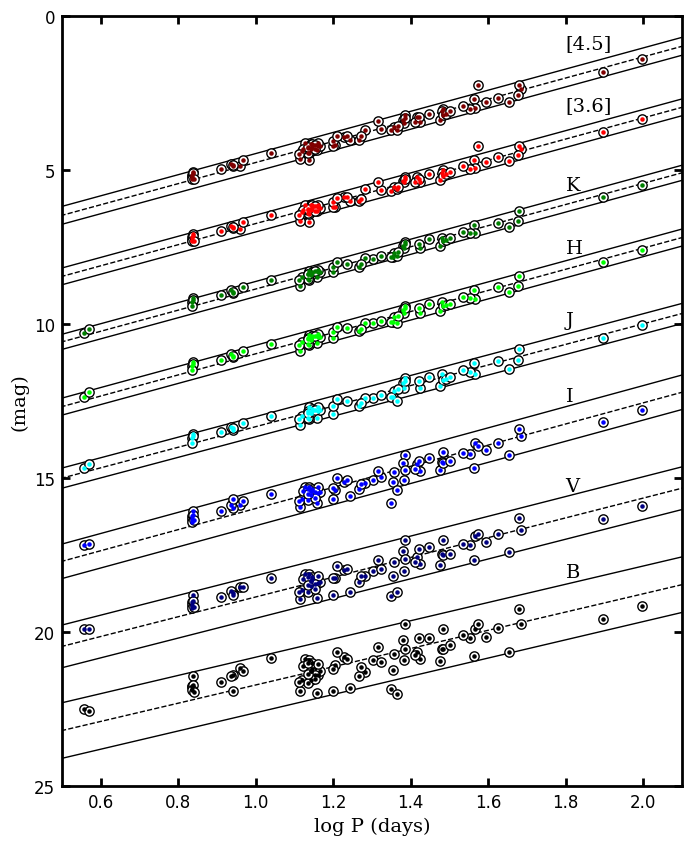}
    \caption{The time-averaged Period-Luminosity (PL) for LMC Cepheid stars across multiple wavelengths are presented. A trend is observed with the dispersion visibly decreasing from optical to infrared wavelengths, as highlighted by the solid black lines flanking the dashed fitted lines. The solid lines, representing the $2\sigma$ limits, indicate the intrinsic width of the PL relation arising from the instability strip. The slopes and intercepts of the fitted lines were adopted from \cite{Madore_2011_multi}.}
    \label{fig:PLR_filters_ALLinONE}
\end{figure}
\vspace{1cm}
\section{Data} \label{sec:data}
In this study, the primary dataset comprises near-infrared observations adopted from \cite{Persson_2004_nearIR}, featuring 92 LMC Cepheid stars located in low-crowding fields. This dataset is considered robust due to its extensive coverage of luminosities and periods (ranging from 3 to 100 days), as well as its high photometric precision, with typical uncertainties of $0.02$–$0.05$ mag in the filter bands, which serve as good indicators of the quality and accuracy of the photometry. To enhance the analysis, the dataset was supplemented with additional observations in optical (Monson et al. 2025, in preparation) and mid-infrared bands \citep{Scowcroft_2011_Spitzer}. The resulting multiwavelength data, provided as time series with dense phase coverage, ensures high-quality sampling of the stars' variability. Figure~\ref{fig:Single_All_filters} illustrates the phased light curves of two sample stars across the available wavelengths. The multiwavelength dataset is constructed from observations detailed as follows.

\textbf{Near-IR bands (JHK)}: The near-IR data were observed in three J, H and K filter bands using the 1-m Swope and 2.5-m duPont telescopes at Las Campanas Observatory from October 1993 to January 1997 and as published by \cite{Persson_2004_nearIR}. Some of these targets were Cepheids from \cite{laney_1986_infrared}, which were re-observed to investigate potential systematic differences that might have arisen in the course of  advancements in the IR detector technology. These advancements had reduced systematic errors to below the 0.1 mag level \citep{Persson_2004_nearIR}.

\textbf{Mid-IR bands ([3.6][4.5])}: The mid-IR data were collected by the \textit{Spitzer} Space Telescope between October 3, 2009, and July 18, 2010 across 24 epochs of observation, in filters 3.6 $\mu m$ and 4.5 $\mu m$, referred to as IRAC1 and IRAC2. The dataset consists of 85 Cepheids with periods longer than about 6 days, drawn from the \cite{Persson_2004_nearIR} sample, and reported in \cite{Scowcroft_2011_Spitzer}.

\begin{figure*}
    \centering
    \begin{minipage}{1\textwidth}
        \begin{minipage}[t]{0.49\linewidth}
            \centering
            \includegraphics[width=\linewidth]{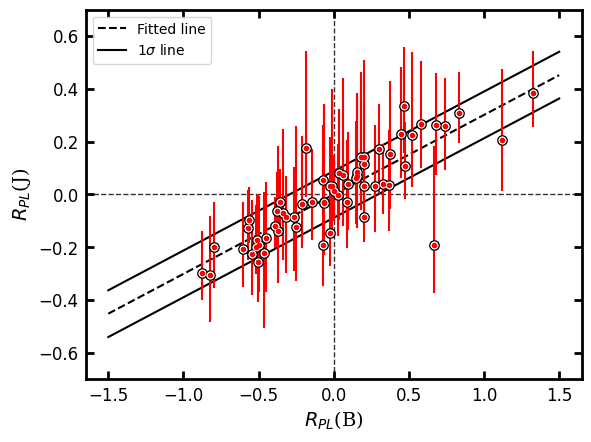}
        \end{minipage}
        \hfill
        \begin{minipage}[t]{0.49\linewidth}
            \centering
            \includegraphics[width=\linewidth]{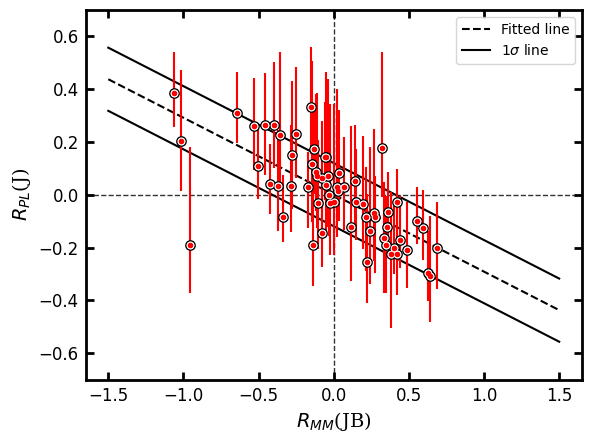}
        \end{minipage}
    \end{minipage}
    \caption{The left panel presents the scatter plot of residuals derived from the PL relation in the $J$ and $B$ filter bands. The right panel depicts the residuals from the fitted lines of the period-magnitude and magnitude-magnitude relations. Both plots exhibit linear relationships, with slopes calculated using a gradient descent algorithm. The error bars indicate errors introduced by the addition of random-phase.}
    \label{fig:combined_residuals}
\end{figure*}

\textbf{Optical bands (BVI)}: The optical data were recently taken by the {\bf T}hree-hundred {\bf M}illi{\bf M}eter {\bf T}elescope (TMMT) located in Chile, between October 27, 2021, and January 30, 2023 (Monson et al. 2025, in preparation), comprising B, V, and I bands. TMMT is a robotic telescope that offers the opportunity to capture modern, high-cadence optical light curves for variable stars, especially those with significant variations like Cepheids (Monson et al. 2025, in preparation), and very short periods, such as less than one day for RR Lyrae \citep{Monson_2017_RRL}.

In general, all the aforementioned datasets consist of time-series observations collected over specific periods. To analyze these data, phased light curves were constructed using the folding method implemented in the \texttt{Lightkurve} package \citep{LightKurve}, adopting periods from Persson et al. (2004), which are themselves based on \citet{Martin_1979_BVI} and references therein.The uncertainties in these periods typically range from $10^{-3}$ to $10^{-2}$ days, which is sufficient for accurate phase alignment over the timescales considered in this study. Examples of the resulting phased light curves are shown as filled circle symbols in Figure~\ref{fig:Single_All_filters}.

The subsequent stage of the analysis involves calculating the time-averaged magnitudes for each star. This process requires observations that are evenly distributed across different phases to ensure uniform phase coverage. However, because the phase distribution in the datasets is non-uniform, the folded light curves cannot be directly used to determine the time-averaged magnitudes.

The next step involves the extraction of uniformly sampled phase points, enabling an unbiased determination of the intensity-averaged, mean magnitude in each selected filter band. For this purpose, the GLOESS method (detailed in Appendix~\ref{sec:GLOESS}) was applied to each star's folded data across all bands, as represented by the solid black dots in Figure~\ref{fig:Single_All_filters}. 

An examination of the light curves in Figure~\ref{fig:Single_All_filters} reveals a transition in their shape from a classic triangular form in the optical bands to a more sinusoidal (or cycloidal) profile in the mid-infrared, accompanied by a corresponding decrease in amplitude \citep{wisniewski_johnson_1968}. This change is primarily due to the diminished influence of temperature variations on surface brightness at longer wavelengths \citep{Freedman_2010_multi_LC} revealing the achromatic radius variations.

Figure~\ref{fig:PLR_filters_ALLinONE} presents the PL relation derived using mean magnitudes of LMC Cepheids computed via the GLOESS method.  As shown, the PL relations span a broad range of wavelengths, from optical to mid-infrared bands. A key observation is the progressive narrowing of the PL relation's dispersion with increasing wavelength. This trend suggests that longer wavelengths, particularly in the mid-infrared, provide a more stable and precise framework for PL relation analysis, likely due to reduced sensitivity to stellar temperature fluctuations  \citep{Freedman_2010_multi_LC} and interstellar extinction \citep{Freedman_1990_Manifold,Persson_2004_nearIR}.

\vspace{0.5cm}

\section{Methodology} \label{sec:methodology}

As previously mentioned, single-epoch observations produce larger dispersion in their PL relations compared to multi-epoch data, due to the random sampling of the light curves. In this section, we describe  in detail the methodology used to mitigate the PL dispersion resulting from the use of single-epoch data. The approach starts with an established PL relation in one filter, followed by a series of steps to estimate corrections for random-phase sampling in different filter bands. These steps will be detailed in the following order:

\textbf{Step 1:} Calculating the magnitude residuals in the respective Period-Luminosity relations: The residual represents the deviation of an individual star from the mean/fitted PL relation, as indicated by the dashed line in Figure~\ref{fig:PLR_filters_ALLinONE}. It is defined as follows: 
\begin{equation}\label{eq:RPL} 
R_{PL}(i,j) = M(i,j) - (\alpha_{PL}(i) \times  log[P(j)] + \beta_{PL}(i))
\end{equation}
For each of the $i \in [1,m]$ bands in which time-averaged period-luminosity relations are available (i.e., from B to 4.5$\mu m$ filters, in our case), consider the   $j \in [1,N]$  Cepheid variables with observed periods, $P(j)$ (in days), and time-averaged magnitudes, $M(i,j)$. The two coefficients, $\alpha_{PL}$ and $\beta_{PL}$, are the slope and intercept, respectively, obtained from the PL fit (dashed lines in Figure~\ref{fig:PLR_filters_ALLinONE}). 

The period $P(j)$ is adopted from \citet{Persson_2004_nearIR}, who compiled periods from literature available at the time of their initial observations, particularly from \cite{Martin_1979_BVI} and references therein. The time-averaged magnitude $M(i,j)$ calculated as the mean magnitude obtained after applying the GLOESS method, as described in Appendix~\ref{sec:GLOESS}. This method generates a single-cycle light curve with uniform phases, ensuring that data points are consistently spaced in phase space. 

It is important to note that the PL relation, up to this step, is derived using the time-averaged magnitude $M(i,j)$. Therefore, if a star's position lies above or below the fitted line in one filter band, it will exhibit the same offset direction in all other bands \citep{Madore_2024_NovelMethod}. 

\textbf{Step 2:} Calculating the residuals in the magnitude-magnitude space: As shown in Equation \ref{eq:RMM}, this residual quantifies the signed difference between a star's position on the magnitude-magnitude diagram and the fitted line on this plane. Figure~\ref{fig:AA_all} illustrates all possible magnitude-magnitude diagrams in our dataset. The solid line in each subplot represents the best-fit of distributions in the data and serves as the reference for calculating residuals.

\begin{figure*}
	\begin{center}
	\begin{minipage}{.99\textwidth}
		\includegraphics[width=0.325\linewidth]{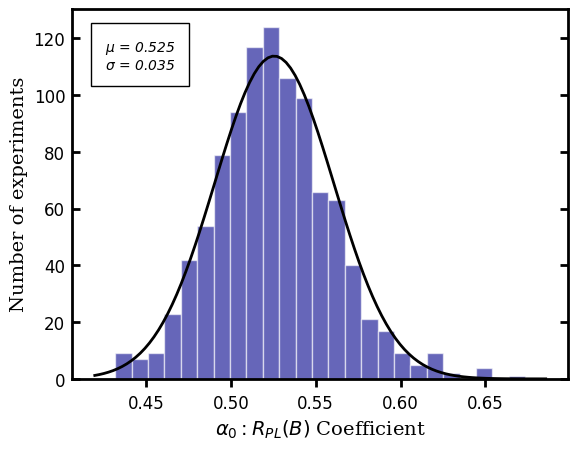}
		\includegraphics[width=0.33\linewidth]{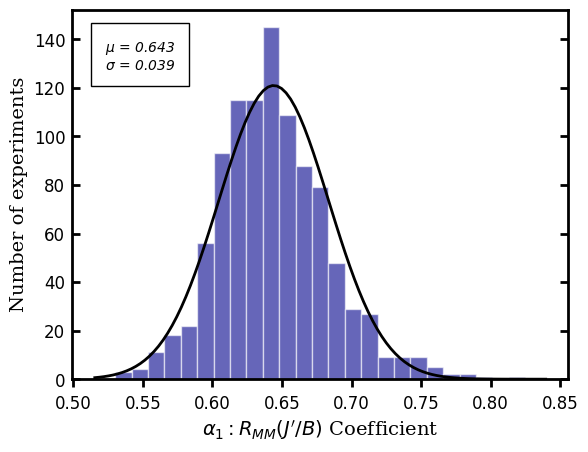}
            \includegraphics[width=0.325\linewidth]{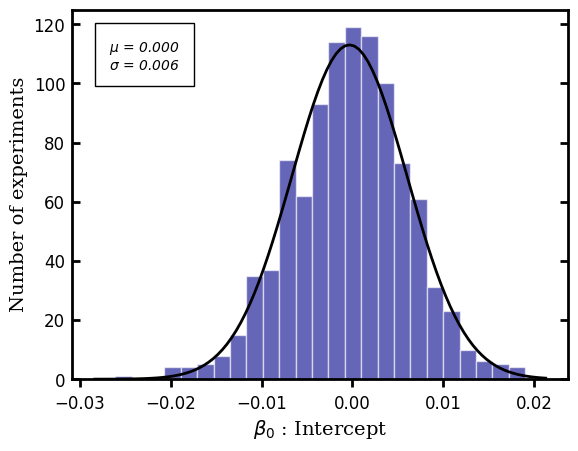}
	\end{minipage}
        \caption{From Left to Right Panels: Distributions of the coefficients and intercept derived from Equation~\ref{eq:correction_raw} across 1000 experiments using random-phase data. All distributions exhibit Gaussian behavior, allowing for the precise determination of the coefficients and intercept.}
        \label{fig:Par_cor}
	\end{center}
\end{figure*}

\begin{equation}\label{eq:RMM} 
R_{MM}(ik,j) = M(i,j) - (\alpha_{MM} \times  M(k,j) - \beta_{MM})
\end{equation}
where $\alpha_{MM}$ and $\beta_{MM}$ are the slope and intercept obtained from the fitted line (solid lines in Figure~\ref{fig:Pairplot_AA}). $M(k,j)$ represents the time-averaged magnitude for any filter other than filter $i$.

\textbf{Step 3:} Considering Equation \ref{eq:RMM}, it is now feasible to perform the calculations using the $k$ band in order to determine the corrections for random-phase effects in the $i$ filter band. By applying Equations~\ref{eq:RPL} and \ref{eq:RMM}, we obtain:

\begin{equation}\label{eq:correlation_raw}
\begin{split}
R_{PL}(i,j) &\propto    {R_{PL}(k,j)}\\
R_{PL}(i,j) &\propto    {R_{MM}(ik,j)}\\
\end{split}
\end{equation}

Quantifying Equation~\ref{eq:correlation_raw} in more detail leads to:
\begin{equation}\label{eq:RPLCorr}
\begin{split}
R_{PL}(i,j) & = a_{0} \times {R_{PL}(k,j)} + b_{0}\\
R_{PL}(i,j) & = a_{1} \times {R_{MM}(ik,j)} + b_{1}\\
\end{split}
\end{equation}
that can be merged as following equation:
\begin{equation}\label{eq:RPLV0}
\begin{split}
R_{PL}(i,j) = (a_{0}/2) \times {R_{PL}(k,j)} \\ +  (a_{1}/2) \times  {R_{MM}(ik,j)} \\ + (b_{0}+b_{1})/2.&\\
\end{split}
\end{equation}

\textbf{Step 4:} In the previous steps, time-averaged magnitudes were utilized. At this stage, it is assumed that the average magnitude in filter $i$ is not available. Under this assumption, and to investigate the influence of random-phase effects on the Equation \ref{eq:RPLV0}, a restorative correction is applied to the residuals: 

\begin{equation}\label{eq:delta_m}
\begin{split}
R_{PL}(i',j) &= R_{PL}(i,j) + \delta{m(i',j)} \\
R_{MM}(i'k,j) &= R_{MM}(ik,j) + \delta{m(i',j)}.\\
\end{split}
\end{equation}

In Equation~\ref{eq:delta_m}, $\delta{m(i',j)}$ denotes a positive or negative deviation from the time-averaged magnitude. As a result, the calculations are adjusted based on this additional parameter. The prime notation is used to indicate the random-phase magnitude derived from this modification for the $i$ band.

\textbf{Step 5:} By substituting components of Equation~\ref{eq:RPLV0} with the corresponding terms from Equations~\ref{eq:RPLCorr} and \ref{eq:delta_m}, the following relation is obtained:

\begin{equation}
\delta{m(i',j)} = [\frac{-a_{0}}{a_{1}}] \times {R_{PL}(k,j)} + {R_{MM}(i'k,j)}+[\frac{b_{0}-b_{1}}{a_{1}}]
\end{equation}

This equation establishes the relationship between the deviation from the time-averaged magnitude and the mentioned residuals. The parameters $a_{0}$, $a_{1}$, $b_{0}$, and $b_{1}$ are free parameters that can be determined through statistical regressions.

\begin{equation}\label{eq:final_delta_m}
\begin{split}
\delta{m(i',j)} &= \alpha_{0} \times  {R_{PL}(k,j)} + \alpha_{1} \times {R_{MM}(i'k,j)} + \beta_{0}
\end{split}
\end{equation}

Our objective is to train a linear regression model for the determination of coefficients ($\alpha_{0}$, $\alpha_{1}$) and the intercept ($\beta_{0}$).

This model is designed to predict the difference between random-phase data and time-averaged data, denoted as $\delta{m(i',j)}$, which are hereafter referred to as the ``random-phase correction.'' To achieve this, the random-phase correction is subtracted from each star individually, resulting in a corrected magnitude, represented as $cm(i', j)$. This process is mathematically expressed as:

\begin{equation} \label{eq:correction_raw}
\begin{split}
cm(i',j) &= m(i',j) - \delta{m(i',j)}\\
&= m(i',j) - \alpha_{0} \times {R_{PL}(k,j)} \\
& \hspace{11ex} - \alpha_{1} \times {R_{MM}(i'k,j)}\\
& \hspace{11ex} - \beta_{0}\\
\end{split}
\end{equation}

By following the steps outlined in this section, the added scatter produced by random-phase sampling of a given bandpass by a single-epoch observation can be reduced using time-averaged data from another filter. Figures~\ref{fig:Pairplot_RR} and \ref{fig:Pairplot_AA} illustrate the calculated results for all $i$ and $k$ filter bands.

Since the primary objective of this study is to explore applications that leverage the advantages of infrared space telescopes, the J filter was selected as the i filter for single-epoch observations. Figure 3 (left panel) displays a scatter plot of residuals derived from the PL relations in the $J$ and $B$ bands, while the right panel shows residuals from the fitted period–magnitude and magnitude–magnitude relations. Both panels reveal linear relationships, with slopes determined using a gradient descent algorithm. The figure demonstrates a well-defined trend with acceptable dispersion, accounting for potential variations in the PL residuals due to random-phase observations. Additional combinations of filters beyond $J$ and $B$ are shown in Figures A.2 and A.3.

Consequently, the $B$ band is chosen as the $k$ filter and serves as the secondary band for random-phase correction. However, these choices are not unique, and the method can be applied to other filter combinations. The details of the calculations and the results for these selected filter bands are presented in Section~\ref{sec:results}.


\section{Results}\label{sec:results}

This method relies on the physically-understood correlation between residuals from the fitted PL relation and the magnitude-magnitude relations across different bands, as demonstrated for all available filters in Figures~\ref{fig:Pairplot_RR} and \ref{fig:Pairplot_AA}, and tabulated in Tables \ref{tab:correction_par_1} \& \ref{tab:correction_par_2}. 

\begin{figure*}
        \centering
        \includegraphics[width=0.6\linewidth]{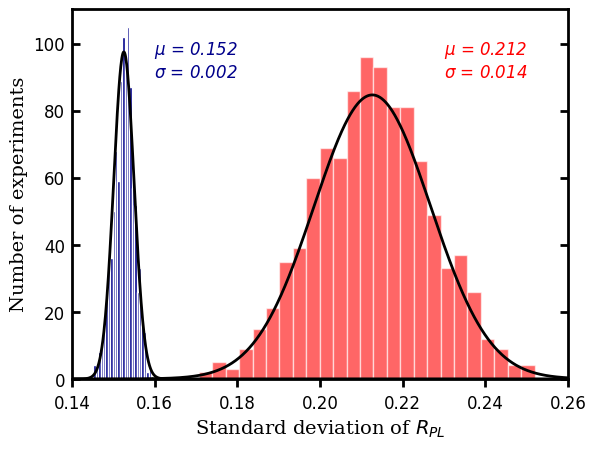}
        \caption{Distribution of the standard deviation of residuals from the PL-fitted line before (red) and after (blue) applying the random-phase correction. As expected with a sufficiently large sample size, both distributions follow a Gaussian shape. The standard deviation decreases from $0.212$ mag to $0.152$ mag after the correction.}
        \label{fig:Std_PLR}
\end{figure*} 

We focus on two filters: one from the NIR ($J$ band), used as single-epoch data, and the other from the optical ($B$ band), employed as time-averaged data for correction. The correlation between these bands, which is used for the correction, is illustrated in Figure~\ref{fig:combined_residuals}.

To calibrate the method described in Section~\ref{sec:methodology} for the selected filters, the experiments were repeated 1000 times, with J magnitudes randomly drawn from the magnitude points generated by the GLOESS method, as illustrated by the filled dots in Figure~\ref{fig:Single_All_filters}, to simulate single-epoch/random-phase data. The variation that each star may experience in an experiment is shown in Figures~\ref{fig:combined_residuals}, where the magnitude variations are approximately half of the magnitude in the $J$ filter band. For the $B$ band, used to derive the random-phase correction, time-averaged data is obtained by averaging the magnitudes generated by the GLOESS method, ensuring a uniformly phased light curve.

In each experiment, the free parameters defined in Equation~\ref{eq:correction_raw} were optimized using linear regression, with the J-band random-phase data corrected based on the time-averaged B-band magnitudes. This process resulted in 1000 values for each free parameter, which distributions are illustrated in Figure~\ref{fig:Par_cor}. By performing statistical analysis, including Gaussian fitting to these distributions, we derive the coefficients and the intercept along with their associated 1$\sigma$ uncertainties:

\begin{equation} \label{eq:val}
    \begin{split}
    \alpha_{0} &= 0.525 \pm 0.035 \\
    \alpha_{1} &= 0.643 \pm 0.039 \\
    \beta_{0} &= 0.000 \pm 0.006
    \end{split}
\end{equation}

The effectiveness of this method can be evaluated by measuring the standard deviation of the residuals from the fitted PL relation. This standard deviation quantifies the dispersion in the PL relation and serves as a key metric for evaluating the method’s precision. Figure~\ref{fig:Std_PLR} shows the distribution of this parameter before and after the random-phase correction. Statistical analysis, which includes modeling the distributions with a Gaussian function, reveals that the mean scatter decreases from $0.212$ to $0.152$ mag, highlighting a reduction in the width of the PL relation after applying the correction to the random-phase J band data.

\begin{figure*}
    \centering
    \includegraphics[width=0.75\linewidth]{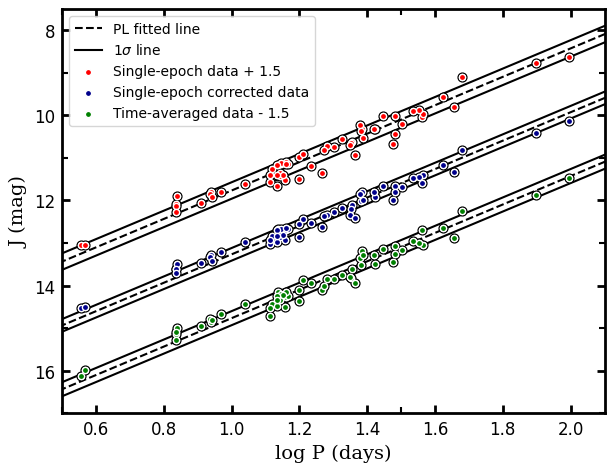}
    \caption{PL relations in the J band shown for single-epoch, time-averaged, and corrected data. The scatter of the PL relations decreases, demonstrating the transition from random-phase data to corrected data, which closely approximates time-averaged data. The dashed lines represent the fitted line using the slope provided by \cite{Madore_2011_multi} and the trained intercepts derived in this work. The solid lines indicate a $1\sigma$ dispersion around the dashed line.}
    \label{fig:RP_CD_J}
\end{figure*}

By substituting the derived coefficients and intercept into Equation~\ref{eq:correction_raw}, a relationship is established that facilitates the correction of magnitudes for stars in the J band obtained from single-epoch/random-phase data. Applying this correction to a random-phase dataset results in a narrower PL relation compared to uncorrected random-phase PL data, as shown in Figure~\ref{fig:RP_CD_J}. Specifically, the scatter of the PL relation increases from $0.195$ mag in time-averaged data to $0.212$ mag in random-phase data. However, after applying the correction, the width decreases by approximately 1.4 times, reducing the width to $0.152$ mag, as illustrated in Figure~\ref{fig:Std_PLR}. This reduction corresponds to a $28\%$ decrease in the statistical uncertainty of the apparent distance modulus, calculated as $(0.212 - 0.152) / 0.212 = 0.28$.

One might ask how many observations would be needed to be added to the single J-band observation in order to obtain the same precision as delivered by the magnitude-magnitude relation.
We first note that the scatter in mean-light PL relations is very similar in magnitude to the scatter added by the random sampling of the light curves in the same band. In other words the variance in the random-phase PL relation is composed of two terms each of which has the same size. For instance, in the example given above, if the scatter in the J band introduced by a randomly-sampled is $\sigma_{random} = \pm$ 0.15~mag and the scatter in the time-averaged J band PL relation is also $\sigma_{mean} = \pm$ 0.15 mag then the predicted scatter in the random-phase J-band PL relation would be those two sources of scatter added in quadrature yielding $\sqrt{(0.15)^2 +(0.15)^2} =$ 0.21~mag, as indeed it is seen to be, above.
Bringing down the scatter in the observed PL relation by increasing the number of epochs at a fixed number of Cepheids would take on the form $\sigma _{N,total} = \sqrt{(0.15/N)^2 + (0.15)^2}$, which for N = 10 gives  $\sigma_{10, total} = 0.16 $ mag. Without over-selling the power of the method discussed in this paper, it is clear that {\bf at least} an order of magnitude fewer observations can reduce the observed scatter compared to the same reduction obtained by significantly increased numbers of observations/epochs.

\section{conclusion} \label{sec:conlusion}

In this study, we expand and elaborate on a method to correct the errors inherent in single-epoch/random-phase data, bringing it closer to the time-averaged data required to construct accurate PL relations.

This correction is based on the correlation between two distinct residuals: the residuals from the time-averaged PL relations, and the residuals from the magnitude-magnitude relation between two filters. While this approach is applicable to any combination of filters, we specifically validate the method using J-band single-epoch magnitudes and B-band time-averaged magnitudes. By analyzing observations of stars in the LMC and conducting 1000 experiments on them, we statistically determine the free parameters of the correction equation.

This correction statistically reduces the width of the PL relation by a factor of 1.4, corresponding to a $28\%$ decrease in statistical uncertainties. The applicability of this method can be further explored for galaxies with diverse properties, such as metallicity, which remains a topic of ongoing debate.  One of the key advantages of this method is its ability to optimize data from telescopes that provide single-epoch observations in one filter (e.g., a near-infrared band), for stars whose time-averaged magnitudes are known in a different filter (e.g., an optical band), assuming well-determined periods and phases. This is particularly useful for observations with the \textit{James Webb} Space Telescope (\textit{JWST}). By reducing phase-related uncertainties, this approach enables more precise photometric measurements of distant galaxies. These improvements enhance the quoted precision of PL relations in such systems, leading to more efficient and precise application of Cepheids to the cosmic distances.


\bibliographystyle{aasjournal}
\bibliography{sample631}{}

\begin{thebibliography}{}
\expandafter\ifx\csname natexlab\endcsname\relax\def\natexlab#1{#1}\fi
\providecommand{\url}[1]{\href{#1}{#1}}
\providecommand{\dodoi}[1]{doi:~\href{http://doi.org/#1}{\nolinkurl{#1}}}
\providecommand{\doeprint}[1]{\href{http://ascl.net/#1}{\nolinkurl{http://ascl.net/#1}}}
\providecommand{\doarXiv}[1]{\href{https://arxiv.org/abs/#1}{\nolinkurl{https://arxiv.org/abs/#1}}}

\bibitem[{Alcock {et~al.}(1997)Alcock, Allsman, Alves, Axelrod, Becker,
  Bennett, Cook, Freeman, Griest, Guern, Lehner, Marshall, Peterson, Pratt,
  Quinn, Rodgers, Stubbs, Sutherland, Welch, \&
  Collaboration)}]{Alcock_1997_MACHO}
Alcock, C., Allsman, R.~A., Alves, D., {et~al.} 1997, The Astrophysical
  Journal, 486, 697, \dodoi{10.1086/304535}

\bibitem[{{Anderson} \& {Riess}(2018)}]{Anderson_2018}
{Anderson}, R.~I., \& {Riess}, A.~G. 2018, apj, 861, 36,
  \dodoi{10.3847/1538-4357/aac5e2}

\bibitem[{{Bhardwaj, A.} {et~al.}(2024){Bhardwaj, A.}, {Ripepi, V.}, {Testa,
  V.}, {Molinaro, R.}, {Marconi, M.}, {De Somma, G.}, {Trentin, E.}, {Musella,
  I.}, {Storm, J.}, {Sicignano, T.}, \& {Catanzaro,
  G.}}]{Bhardwaj_2024_MultiW_CMetall}
{Bhardwaj, A.}, {Ripepi, V.}, {Testa, V.}, {et~al.} 2024, A\&A, 683, A234,
  \dodoi{10.1051/0004-6361/202348140}

\bibitem[{Breuval {et~al.}(2022)Breuval, Riess, Kervella, Anderson, \&
  Romaniello}]{Breuval_2022_Metallicity}
Breuval, L., Riess, A.~G., Kervella, P., Anderson, R.~I., \& Romaniello, M.
  2022, The Astrophysical Journal, 939, 89, \dodoi{10.3847/1538-4357/ac97e2}

\bibitem[{{Cioni} {et~al.}(2011){Cioni}, {Clementini}, {Girardi}, {Guandalini},
  {Gullieuszik}, {Miszalski}, {Moretti}, {Ripepi}, {Rubele}, {Bagheri},
  {Bekki}, {Cross}, {de Blok}, {de Grijs}, {Emerson}, {Evans}, {Gibson},
  {Gonzales-Solares}, {Groenewegen}, {Irwin}, {Ivanov}, {Lewis}, {Marconi},
  {Marquette}, {Mastropietro}, {Moore}, {Napiwotzki}, {Naylor}, {Oliveira},
  {Read}, {Sutorius}, {van Loon}, {Wilkinson}, \& {Wood}}]{VISTA_2011}
{Cioni}, M.~R., {Clementini}, G., {Girardi}, L., {et~al.} 2011, The Messenger,
  144, 25

\bibitem[{Cleveland \& Grosse(1991)}]{cleveland_1991_computational}
Cleveland, W.~S., \& Grosse, E. 1991, Statistics and computing, 1, 47

\bibitem[{{Fouqu{\'e}} {et~al.}(2007){Fouqu{\'e}}, {Arriagada}, {Storm},
  {Barnes}, {Nardetto}, {M{\'e}rand}, {Kervella}, {Gieren}, {Bersier},
  {Benedict}, \& {McArthur}}]{Fouqu_2007_MW_data}
{Fouqu{\'e}}, P., {Arriagada}, P., {Storm}, J., {et~al.} 2007, \aap, 476, 73,
  \dodoi{10.1051/0004-6361:20078187}

\bibitem[{{Freedman}(1988)}]{Freedman88}
{Freedman}, W.~L. 1988, apj, 326, 691, \dodoi{10.1086/166128}

\bibitem[{{Freedman} \& {Madore}(1990)}]{Freedman_1990_Manifold}
{Freedman}, W.~L., \& {Madore}, B.~F. 1990, \apj, 365, 186,
  \dodoi{10.1086/169469}

\bibitem[{Freedman \& Madore(2010)}]{Freedman_2010_multi_LC}
Freedman, W.~L., \& Madore, B.~F. 2010, The Astrophysical Journal, 719, 335,
  \dodoi{10.1088/0004-637X/719/1/335}

\bibitem[{{Freedman} \& {Madore}(2024)}]{Freedman_2024_IAUS376}
{Freedman}, W.~L., \& {Madore}, B.~F. 2024, in IAU Symposium, Vol. 376, IAU
  Symposium, ed. R.~{de Grijs}, P.~A. {Whitelock}, \& M.~{Catelan}, 1--14,
  \dodoi{10.1017/S1743921323003459}

\bibitem[{Freedman {et~al.}(2008)Freedman, Madore, Rigby, Persson, \&
  Sturch}]{Freedman_2008_First_Epoch}
Freedman, W.~L., Madore, B.~F., Rigby, J., Persson, S.~E., \& Sturch, L. 2008,
  The Astrophysical Journal, 679, 71, \dodoi{10.1086/586701}

\bibitem[{{Freedman} {et~al.}(2012){Freedman}, {Madore}, {Scowcroft}, {Burns},
  {Monson}, {Persson}, {Seibert}, \& {Rigby}}]{freedman_2012_carnegie}
{Freedman}, W.~L., {Madore}, B.~F., {Scowcroft}, V., {et~al.} 2012, \apj, 758,
  24, \dodoi{10.1088/0004-637X/758/1/24}

\bibitem[{Gardner {et~al.}(2023)Gardner, Mather, Abbott, Abell, Abernathy,
  Abney, Abraham, Abraham, Abul-Huda, Acton, Adams, Adams, Adler, Adriaensen,
  Aguilar, Ahmed, Ahmed, Ahmed, Albat, Albert, Alberts, Aldridge, Allen, Allen,
  Altenburg, Altunc, Alvarez, Álvarez Márquez, de~Oliveira, Ambrose,
  Anandakrishnan, Andersen, Anderson, Anderson, Anderson, Anderson, Aprea,
  Archer, Arenberg, Argyriou, Arribas, Artigau, Arvai, Atcheson, Atkinson,
  Averbukh, Aymergen, Bacinski, Baggett, Bagnasco, Baker, Balzano, Banks,
  Baran, Barker, Barrett, Barringer, Barto, Bast, Baudoz, Baum, Beatty,
  Beaulieu, Bechtold, Beck, Beddard, Beichman, Bellagama, Bely, Berger,
  Bergeron, Bernier, Bertch, Beskow, Betz, Biagetti, Birkmann, Bjorklund,
  Blackwood, Blazek, Blossfeld, Bluth, Boccaletti, Boegner~Jr, Bohlin, Boia,
  Böker, Bonaventura, Bond, Bosley, Boucarut, Bouchet, Bouwman, Bower, Bowers,
  Bowers, Boyce, Boyer, Boyer, Boyer, Boyer, Bradley, Brady, Brandl, Brannen,
  Breda, Bremmer, Brennan, Bresnahan, Bright, Broiles, Bromenschenkel, Brooks,
  Brooks, Brown, Brown, Brown, Bruce, Bryson, Bujanda, Bullock, Bunker, Bureo,
  Burt, Bush, Bushouse, Bussman, Cabaud, Cale, Calhoon, Calvani, Canipe,
  Caputo, Cara, Carey, Case, Cesari, Cetorelli, Chance, Chandler, Chaney,
  Chapman, Charlot, Chayer, Cheezum, Chen, Chen, Cherinka, Chichester, Chilton,
  Chittiraibalan, Clampin, Clark, Clark, Clark, Claybrooks, Cleveland, Cohen,
  Cohen, Colón, Coleman, Colina, Comber, Comeau, Comer, Reis, Connolly,
  Conroy, Contos, Contreras, Cook, Cooper, Cooper, Correia, Correnti, Cossou,
  Costanza, Coulais, Cox, Coyle, Cracraft, Crew, Curtis, Cusveller, Maciel,
  Dailey, Daugeron, Davidson, Davies, Davis, Davis, Day, de~Chambure, de~Jong,
  De~Marchi, Dean, Decker, Delisa, Dell, Dellagatta, Dembinska, Demosthenes,
  Dencheva, Deneu, DePriest, Deschenes, Dethienne, Detre, Diaz, Dicken,
  DiFelice, Dillman, Disharoon, Dixon, Doggett, Dominguez, Donaldson,
  Doria-Warner, Santos, Doty, Douglas, Doyon, Dressler, Driggers, Driggers,
  Dunn, DuPrie, Dupuis, Durning, Dutta, Earl, Eccleston, Ecobichon, Egami,
  Ehrenwinkler, Eisenhamer, Eisenhower, Eisenstein, El~Hamel, Elie, Elliott,
  Elliott, Engesser, Espinoza, Etienne, Etxaluze, Evans, Fabreguettes,
  Falcolini, Falini, Fatig, Feeney, Feinberg, Fels, Ferdous, Ferguson,
  Ferrarese, Ferreira, Ferruit, Ferry, Filippazzo, Firre, Fix, Flagey,
  Flanagan, Fleming, Florian, Flynn, Foiadelli, Fontaine, Fontanella, Forshay,
  Fortner, Fox, Framarini, Francisco, Franck, Franx, Franz, Friedman, Friend,
  Frost, Fu, Fullerton, Gaillard, Galkin, Gallagher, Galyer, García~Marín,
  Gardner, Garland, Garrett, Gasman, Gáspár, Gastaud, Gaudreau, Gauthier,
  Geers, Geithner, Gennaro, Gerber, Gereau, Giampaoli, Giardino, Gibbons,
  Gilbert, Gilman, Girard, Giuliano, Gkountis, Glasse, Glassmire, Glauser,
  Glazer, Goldberg, Golimowski, Gonzaga, Gordon, Gordon, Goudfrooij, Gough,
  Graham, Grau, Green, Greene, Greene, Greenfield, Greenhouse, Greve, Greville,
  Grimaldi, Groe, Groebner, Grumm, Grundy, Güdel, Guillard, Guldalian, Gunn,
  Gurule, Gutman, Guy, Guyot, Hack, Haderlein, Hagan, Hagedorn, Hainline,
  Haley, Hami, Hamilton, Hammann, Hammel, Hanley, Hansen, Hardy, Harnisch,
  Harr, Harris, Hart, Hartig, Hasan, Hashim, Hashimoto, Haskins, Hawkins,
  Hayden, Hayden, Healy, Hecht, Heeg, Hejal, Helm, Hengemihle, Henning, Henry,
  Henry, Henshaw, Hernandez, Herrington, Heske, Hesman, Hickey, Hilbert, Hines,
  Hinz, Hirsch, Hitcho, Hodapp, Hodge, Hoffman, Holfeltz, Holler, Hoppa,
  Horner, Howard, Howard, Huber, Hunkeler, Hunter, Hunter, Hurd, Hurst,
  Hutchings, Hylan, Ignat, Illingworth, Irish, Isaacs~III, Jackson~Jr, Jaffe,
  Jahic, Jahromi, Jakobsen, James, James, James, Jamieson, Jandra,
  Jayawardhana, Jedrzejewski, Jeffers, Jensen, Joanne, Johns, Johnson, Johnson,
  Johnson, Johnson, Johnson, Johnson, Johnstone, Jollet, Jones, Jones, Jones,
  Jones, Jones, Jordan, Jordan, Jue, Jurkowski, Justis, Justtanont, Kaleida,
  Kalirai, Kalmanson, Kaltenegger, Kammerer, Kan, Kanarek, Kao, Karakla, Karl,
  Kassin, Kauffman, Kavanagh, Kelley, Kelly, Kendrew, Kennedy, Kenny,
  Keski-Kuha, Keyes, Khan, Kidwell, Kimble, King, King, Kinzel, Kirk,
  Kirkpatrick, Klaassen, Klingemann, Klintworth, Knapp, Knight, Knollenberg,
  Knutsen, Koehler, Koekemoer, Kofler, Kontson, Kovacs, Kozhurina-Platais,
  Krause, Kriss, Krist, Kristoffersen, Krogel, Krueger, Kulp, Kumari, Kwan,
  Kyprianou, Labador, Labiano, Lafrenière, Lagage, Laidler, Laine, Laird,
  Lajoie, Lallo, Lam, LaMassa, Lambros, Lampenfield, Lander, Langston, Larson,
  Larson, LaVerghetta, Law, Lawrence, Lee, Lee, Lee, Leisenring, Leveille,
  Levenson, Levi, Levine, Lewis, Lewis, Lewis, Libralato, Lidon, Liebrecht,
  Lightsey, Lilly, Lim, Lim, Ling, Link, Link, Lipinski, Liu, Lo, Lobmeyer,
  Logue, Long, Long, Long, Long, López-Caniego, Lotz, Love-Pruitt, Lubskiy,
  Luers, Luetgens, Luevano, G.~Flores~Lui, Lund~III, Lundquist, Lunine,
  Lützgendorf, Lynch, MacDonald, MacDonald, Macias, Macklis, Maghami,
  Maharaja, Maiolino, Makrygiannis, Malla, Malumuth, Manjavacas, Marini,
  Marrione, Marston, Martel, Martin, Martin, Martinez, Maschmann, Masci,
  Masetti, Maszkiewicz, Matthews, Matuskey, McBrayer, McCarthy, McCaughrean,
  McClare, McClare, McCloskey, McClurg, McCoy, McElwain, McGregor, McGuffey,
  McKay, McKenzie, McLean, McMaster, McNeil, De~Meester, Mehalick, Meixner,
  Meléndez, Menzel, Menzel, Merz, Mesterharm, Meyer, Meyett, Meza, Midwinter,
  Milam, Miller, Miller, Miskey, Misselt, Mitchell, Mohan, Montoya, Moran,
  Morishita, Moro-Martín, Morrison, Morrison, Morse, Moschos, Moseley, Mosier,
  Mosner, Mountain, Muckenthaler, Mueller, Mueller, Muhiem, Mühlmann,
  Mullally, Mullen, Munger, Murphy, Murray, Muzerolle, Mycroft, Myers, Myers,
  R.~Myers, Myers, Myrick, Nagle, Nayak, Naylor, Neff, Nelan, Nella, Nguyen,
  Nguyen, Nickson, Nidhiry, Niedner, Nieto-Santisteban, Nikolov, Nishisaka,
  Noriega-Crespo, Nota, O’Mara, Oboryshko, O’Brien, Ochs, Offenberg, Ogle,
  Ohl, Olmsted, Osborne, O’Shaughnessy, Östlin, O’Sullivan, Otor, Ottens,
  Ouellette, Outlaw, Owens, Pacifici, Page, Paranilam, Park, Parrish, Paschal,
  Patapis, Patel, Patrick, Pattishall~Jr, Paul, Paul, Pauly, Pavlovsky,
  Peña-Guerrero, Pedder, Peek, Pelham, Penanen, Perriello, Perrin, Perrine,
  Perrygo, Peslier, Petach, Peterson, Pfarr, Pierson, Pietraszkiewicz, Pilchen,
  Pipher, Pirzkal, Pitman, Player, Plesha, Plitzke, Pohner, Poletis, Pollizzi,
  Polster, Pontius, Pontoppidan, Porges, Potter, Prescott, Proffitt, Pueyo,
  Quispe~Neira, Radich, Rager, Rameau, Ramey, Alarcon, Rampini, Rapp, Rashford,
  Rauscher, Ravindranath, Rawle, Rawlings, Ray, Regan, Rehm, Rehm, Reid, Reis,
  Renk, Reoch, Ressler, Rest, Reynolds, Richon, Richon, Ridgaway, Riedel,
  Rieke, Rieke, Rifelli, Rigby, Riggs, Ringel, Ritchie, Rix, Robberto,
  Robinson, Robinson, Robinson, Rock, Rodriguez, del Pino, Roellig, Rohrbach,
  Roman, Romelfanger, Romo~Jr, Rosales, Rose, Roteliuk, Roth, Rothwell,
  Rouzaud, Rowe, Rowlands, Roy, Royer, Rui, Rumler, Rumpl, Russ, Ryan, Ryan,
  Saad, Sabata, Sabatino, Sabbi, Sabelhaus, Sabia, Sahu, Saif, Salvignol,
  Samara-Ratna, Samuelson, Sanders, Sappington, Sargent, Sauer, Savadkin,
  Sawicki, Schappell, Scheffer, Scheithauer, Scherer, Schiff, Schlawin,
  Schmeitzky, Schmitz, Schmude, Schneider, Schreiber, Schroeven-Deceuninck,
  Schultz, Schwab, Schwartz, Scoccimarro, Scott, Scott, Seaton, Seely, Seery,
  Seidleck, Sembach, Shanahan, Shaughnessy, Shaw, Shay, Sheehan, Sheth, Shih,
  Shivaei, Siegel, Sienkiewicz, Simmons, Simon, Sirianni, Sivaramakrishnan,
  Slade, Sloan, Slocum, Slowinski, Smith, Smith, Smith, Smith, Smith, Smith,
  Smolik, Soderblom, Sohn, Sokol, Sonneborn, Sontag, Sooy, Soummer, Southwood,
  Spain, Sparmo, Speer, Spencer, Sprofera, Stallcup, Stanley, Stansberry,
  Stark, Starr, Stassi, Steck, Steeley, Stephens, Stephenson, Stewart,
  Stiavelli, Jr, Strada, Straughn, Streetman, Strickland, Strobele, Stuhlinger,
  Stys, Such, Sukhatme, Sullivan, Sullivan, Sumner, Sun, Sunnquist, Swade,
  Swam, Swenton, Swoish, Tam~Litten, Tamas, Tao, Taylor, Taylor, Plate,
  Van~Tea, Teague, Telfer, Temim, Texter, Thatte, Thompson, Thompson, Thomson,
  Thronson, Tierney, Tikkanen, Tinnin, Tippet, Todd, Tran, Trauger, Trejo,
  Vinh~Truong, Tsukamoto, Tufail, Tumlinson, Tustain, Tyra, Ubeda, Underwood,
  Uzzo, Vaclavik, Valenduc, Valenti, Van~Campen, van~de Wetering, Van
  Der~Marel, van Haarlem, Vandenbussche, van Dishoeck, Vanterpool, Vernoy,
  Vila~Costas, Volk, Voorzaat, Voyton, Vydra, Waddy, Waelkens, Wahlgren,
  Walker~Jr, Wander, Warfield, Warner, Wasiak, Wasiak, Wehner, Weiler, Weilert,
  Weiss, Wells, Welty, Wheate, Wheeler, White, Whitehouse, Whiteleather,
  Whitman, Williams, Willmer, Willott, Willoughby, Wilson, Wilson, Wilson,
  Windhorst, Wislowski, Wolfe, Wolfe, Wolff, Wondel, Woo, Woods, Worden,
  Workman, Wright, Wu, Wu, Wun, Wymer, Yadetie, Yan, Yang, Yates, Yeager,
  Yerger, Young, Young, Yu, Yu, Zak, Zeidler, Zepp, Zhou, Zincke, Zonak, \&
  Zondag}]{Gardner_2023_JWST}
Gardner, J.~P., Mather, J.~C., Abbott, R., {et~al.} 2023, Publications of the
  Astronomical Society of the Pacific, 135, 068001,
  \dodoi{10.1088/1538-3873/acd1b5}

\bibitem[{Gehrels {et~al.}(2015)Gehrels, Baltay, Bennett, Breckinridge,
  Donahue, Dressler, Gaudi, Greene, Guyon, {et~al.}}]{spergel_2015_RST}
Gehrels, N., Baltay, C., Bennett, D., {et~al.} 2015, arXiv preprint
  arXiv:1503.03757

\bibitem[{{Gibson} {et~al.}(2000{\natexlab{a}}){Gibson}, {Maloney}, \&
  {Sakai}}]{Gibson_2000}
{Gibson}, B.~K., {Maloney}, P.~R., \& {Sakai}, S. 2000{\natexlab{a}}, apjl,
  530, L5, \dodoi{10.1086/312477}

\bibitem[{{Gibson} {et~al.}(2000{\natexlab{b}}){Gibson}, {Stetson}, {Freedman},
  {Mould}, {Kennicutt}, {Huchra}, {Sakai}, {Graham}, {Fassett}, {Kelson},
  {Ferrarese}, {Hughes}, {Illingworth}, {Macri}, {Madore}, {Sebo}, \&
  {Silbermann}}]{Gibson2000}
{Gibson}, B.~K., {Stetson}, P.~B., {Freedman}, W.~L., {et~al.}
  2000{\natexlab{b}}, apj, 529, 723, \dodoi{10.1086/308306}

\bibitem[{{Kimble} {et~al.}(2008){Kimble}, {MacKenty}, {O'Connell}, \&
  {Townsend}}]{Kimble_2008_NIR_HST}
{Kimble}, R.~A., {MacKenty}, J.~W., {O'Connell}, R.~W., \& {Townsend}, J.~A.
  2008, in Society of Photo-Optical Instrumentation Engineers (SPIE) Conference
  Series, Vol. 7010, Space Telescopes and Instrumentation 2008: Optical,
  Infrared, and Millimeter, ed. J.~M. {Oschmann}, Jr., M.~W.~M. {de Graauw}, \&
  H.~A. {MacEwen}, 70101E, \dodoi{10.1117/12.789581}

\bibitem[{{Labhardt} {et~al.}(1997){Labhardt}, {Sandage}, \&
  {Tammann}}]{Labhardt97}
{Labhardt}, L., {Sandage}, A., \& {Tammann}, G.~A. 1997, aap, 322, 751

\bibitem[{Laney \& Stobie(1986)}]{laney_1986_infrared}
Laney, C., \& Stobie, R. 1986, Monthly Notices of the Royal Astronomical
  Society, 222, 449

\bibitem[{{Leavitt} \& {Pickering}(1912)}]{Leavitt_1912_Cep_PLR}
{Leavitt}, H.~S., \& {Pickering}, E.~C. 1912, Harvard College Observatory
  Circular, 173, 1

\bibitem[{{Lightkurve Collaboration} {et~al.}(2018){Lightkurve Collaboration},
  {Cardoso}, {Hedges}, {Gully-Santiago}, {Saunders}, {Cody}, {Barclay}, {Hall},
  {Sagear}, {Turtelboom}, {Zhang}, {Tzanidakis}, {Mighell}, {Coughlin}, {Bell},
  {Berta-Thompson}, {Williams}, {Dotson}, \& {Barentsen}}]{LightKurve}
{Lightkurve Collaboration}, {Cardoso}, J.~V.~d.~M., {Hedges}, C., {et~al.}
  2018, {Lightkurve: Kepler and TESS time series analysis in Python},
  Astrophysics Source Code Library.
\newblock \doeprint{1812.013}

\bibitem[{Macri {et~al.}(2015)Macri, Ngeow, Kanbur, Mahzooni, \&
  Smitka}]{Macri_2015_NIR}
Macri, L.~M., Ngeow, C.-C., Kanbur, S.~M., Mahzooni, S., \& Smitka, M.~T. 2015,
  The Astronomical Journal, 149, 117, \dodoi{10.1088/0004-6256/149/4/117}

\bibitem[{{Madore} \& {Freedman}(1991)}]{Madore_freedman_1991}
{Madore}, B.~F., \& {Freedman}, W.~L. 1991, Publications of the Astronomical
  Society of the Pacific, 103, 933, \dodoi{10.1086/132911}

\bibitem[{Madore \& Freedman(2011)}]{Madore_2011_multi}
Madore, B.~F., \& Freedman, W.~L. 2011, The Astrophysical Journal, 744, 132,
  \dodoi{10.1088/0004-637X/744/2/132}

\bibitem[{Madore {et~al.}(2017)Madore, Freedman, \&
  Moak}]{Madore_2017_EX_DM_Cor}
Madore, B.~F., Freedman, W.~L., \& Moak, S. 2017, The Astrophysical Journal,
  842, 42, \dodoi{10.3847/1538-4357/aa6e4d}

\bibitem[{Madore {et~al.}(2024)Madore, Freedman, \&
  Owens}]{Madore_2024_NovelMethod}
Madore, B.~F., Freedman, W.~L., \& Owens, K. 2024, The Astronomical Journal,
  167, 201, \dodoi{10.3847/1538-3881/ad2c87}

\bibitem[{Madore {et~al.}(2009)Madore, Freedman, Rigby, Persson, Sturch, \&
  Mager}]{Madore_2009_Second_Epoch}
Madore, B.~F., Freedman, W.~L., Rigby, J., {et~al.} 2009, The Astrophysical
  Journal, 695, 988, \dodoi{10.1088/0004-637X/695/2/988}

\bibitem[{{Martin} {et~al.}(1979){Martin}, {Warren}, \&
  {Feast}}]{Martin_1979_BVI}
{Martin}, W.~L., {Warren}, P.~R., \& {Feast}, M.~W. 1979, \mnras, 188, 139,
  \dodoi{10.1093/mnras/188.1.139}

\bibitem[{{McGonegal} {et~al.}(1982){McGonegal}, {McAlary}, {Madore}, \&
  {McLaren}}]{McGonegal_1982_First_NIR_PLR}
{McGonegal}, R., {McAlary}, C.~W., {Madore}, B.~F., \& {McLaren}, R.~A. 1982,
  \apjl, 257, L33, \dodoi{10.1086/183803}

\bibitem[{Meixner {et~al.}(2006)Meixner, Gordon, Indebetouw, Hora, Whitney,
  Blum, Reach, Bernard, Meade, Babler, Engelbracht, For, Misselt, Vijh,
  Leitherer, Cohen, Churchwell, Boulanger, Frogel, Fukui, Gallagher, Gorjian,
  Harris, Kelly, Kawamura, Kim, Latter, Madden, Markwick-Kemper, Mizuno,
  Mizuno, Mould, Nota, Oey, Olsen, Onishi, Paladini, Panagia, Perez-Gonzalez,
  Shibai, Sato, Smith, Staveley-Smith, Tielens, Ueta, Dyk, Volk, Werner, \&
  Zaritsky}]{SAGE_2006}
Meixner, M., Gordon, K.~D., Indebetouw, R., {et~al.} 2006, The Astronomical
  Journal, 132, 2268, \dodoi{10.1086/508185}

\bibitem[{{Monson} {et~al.}(2017){Monson}, {Beaton}, {Scowcroft}, {Freedman},
  {Madore}, {Rich}, {Seibert}, {Kollmeier}, \& {Clementini}}]{Monson_2017_RRL}
{Monson}, A.~J., {Beaton}, R.~L., {Scowcroft}, V., {et~al.} 2017, \aj, 153, 96,
  \dodoi{10.3847/1538-3881/153/3/96}

\bibitem[{{Ngeow} {et~al.}(2003){Ngeow}, {Kanbur}, {Nikolaev}, {Tanvir}, \&
  {Hendry}}]{Ngeow03}
{Ngeow}, C.-C., {Kanbur}, S.~M., {Nikolaev}, S., {Tanvir}, N.~R., \& {Hendry},
  M.~A. 2003, apj, 586, 959, \dodoi{10.1086/367698}

\bibitem[{{Nikolaev} {et~al.}(2004){Nikolaev}, {Drake}, {Keller}, {Cook},
  {Dalal}, {Griest}, {Welch}, \& {Kanbur}}]{Nikolaev04}
{Nikolaev}, S., {Drake}, A.~J., {Keller}, S.~C., {et~al.} 2004, apj, 601, 260,
  \dodoi{10.1086/380439}

\bibitem[{{Persson} {et~al.}(2004){Persson}, {Madore}, {Krzemi{\'n}ski},
  {Freedman}, {Roth}, \& {Murphy}}]{Persson_2004_nearIR}
{Persson}, S.~E., {Madore}, B.~F., {Krzemi{\'n}ski}, W., {et~al.} 2004, \aj,
  128, 2239, \dodoi{10.1086/424934}

\bibitem[{{Pietrzy{\'n}ski} {et~al.}(2019){Pietrzy{\'n}ski}, {Graczyk},
  {Gallenne}, {Gieren}, {Thompson}, {Pilecki}, {Karczmarek}, {G{\'o}rski},
  {Suchomska}, {Taormina}, {Zgirski}, {Wielg{\'o}rski}, {Ko{\l}aczkowski},
  {Konorski}, {Villanova}, {Nardetto}, {Kervella}, {Bresolin}, {Kudritzki},
  {Storm}, {Smolec}, \& {Narloch}}]{Pietrzy_2019_LMC_Cal}
{Pietrzy{\'n}ski}, G., {Graczyk}, D., {Gallenne}, A., {et~al.} 2019, \nat, 567,
  200, \dodoi{10.1038/s41586-019-0999-4}

\bibitem[{{Riess} {et~al.}(2019){Riess}, {Casertano}, {Yuan}, {Macri}, \&
  {Scolnic}}]{Riess_2019_LMC_Cal}
{Riess}, A.~G., {Casertano}, S., {Yuan}, W., {Macri}, L.~M., \& {Scolnic}, D.
  2019, \apj, 876, 85, \dodoi{10.3847/1538-4357/ab1422}

\bibitem[{{Riess} {et~al.}(2023){Riess}, {Anand}, {Yuan}, {Casertano},
  {Dolphin}, {Macri}, {Breuval}, {Scolnic}, {Perrin}, \& {Anderson}}]{Riess23}
{Riess}, A.~G., {Anand}, G.~S., {Yuan}, W., {et~al.} 2023, apjl, 956, L18,
  \dodoi{10.3847/2041-8213/acf769}

\bibitem[{{Riess} {et~al.}(2024){Riess}, {Anand}, {Yuan}, {Casertano},
  {Dolphin}, {Macri}, {Breuval}, {Scolnic}, {Perrin}, \&
  {Anderson}}]{Riess_2024_JWST}
---. 2024, \apjl, 962, L17, \dodoi{10.3847/2041-8213/ad1ddd}

\bibitem[{{Ripepi} {et~al.}(2017){Ripepi}, {Cioni}, {Moretti}, {Marconi},
  {Bekki}, {Clementini}, {de Grijs}, {Emerson}, {Groenewegen}, {Ivanov},
  {Molinaro}, {Muraveva}, {Oliveira}, {Piatti}, {Subramanian}, \& {van
  Loon}}]{Ripepi_2017_NIR_SMC}
{Ripepi}, V., {Cioni}, M.-R.~L., {Moretti}, M.~I., {et~al.} 2017, \mnras, 472,
  808, \dodoi{10.1093/mnras/stx2096}

\bibitem[{Ripepi {et~al.}(2022)Ripepi, Chemin, Molinaro, Cioni, Bekki,
  Clementini, de Grijs, De Somma, El Youssoufi, Girardi, Groenewegen,
  Ivanov, Marconi, McMillan, \& van Loon}]{Ripepi_2022_NIR}
Ripepi, V., Chemin, L., Molinaro, R., {et~al.} 2022, Monthly Notices of the
  Royal Astronomical Society, 512, 563, \dodoi{10.1093/mnras/stac595}

\bibitem[{{Ripepi, V.} {et~al.}(2019){Ripepi, V.}, {Molinaro, R.}, {Musella,
  I.}, {Marconi, M.}, {Leccia, S.}, \& {Eyer,
  L.}}]{Ripepi_2019_Reclassification_Cep_GaiaDR2}
{Ripepi, V.}, {Molinaro, R.}, {Musella, I.}, {et~al.} 2019, A\&A, 625, A14,
  \dodoi{10.1051/0004-6361/201834506}

\bibitem[{{Sandage} \& {Tammann}(1968)}]{Sandage_Tammann_1968_PLR}
{Sandage}, A., \& {Tammann}, G.~A. 1968, \apj, 151, 531, \dodoi{10.1086/149454}

\bibitem[{{Scowcroft} {et~al.}(2016){Scowcroft}, {Freedman}, {Madore},
  {Monson}, {Persson}, {Rich}, {Seibert}, \&
  {Rigby}}]{Scowcroft_2016_Spitzer_SMC}
{Scowcroft}, V., {Freedman}, W.~L., {Madore}, B.~F., {et~al.} 2016, \apj, 816,
  49, \dodoi{10.3847/0004-637X/816/2/49}

\bibitem[{{Scowcroft} {et~al.}(2011){Scowcroft}, {Freedman}, {Madore},
  {Monson}, {Persson}, {Seibert}, {Rigby}, \&
  {Sturch}}]{Scowcroft_2011_Spitzer}
---. 2011, \apj, 743, 76, \dodoi{10.1088/0004-637X/743/1/76}

\bibitem[{{Soszy{\'n}ski} {et~al.}(2005){Soszy{\'n}ski}, {Gieren}, \&
  {Pietrzy{\'n}ski}}]{Soszynski05}
{Soszy{\'n}ski}, I., {Gieren}, W., \& {Pietrzy{\'n}ski}, G. 2005, pasp, 117,
  823, \dodoi{10.1086/431434}

\bibitem[{{Soszy{\'n}ski} {et~al.}(2015){Soszy{\'n}ski}, {Udalski},
  {Szyma{\'n}ski}, {Skowron}, {Pietrzy{\'n}ski}, {Poleski}, {Pietrukowicz},
  {Skowron}, {Mr{\'o}z}, {Koz{\l}owski}, {Wyrzykowski}, {Ulaczyk}, \&
  {Pawlak}}]{Soszy_2015_OGLE}
{Soszy{\'n}ski}, I., {Udalski}, A., {Szyma{\'n}ski}, M.~K., {et~al.} 2015,
  \actaa, 65, 297, \dodoi{10.48550/arXiv.1601.01318}

\bibitem[{{Spergel} {et~al.}(2015){Spergel}, {Gehrels}, {Baltay}, {Bennett},
  {Breckinridge}, {Donahue}, {Dressler}, {Gaudi}, {Greene}, {Guyon}, {Hirata},
  {Kalirai}, {Kasdin}, {Macintosh}, {Moos}, {Perlmutter}, {Postman},
  {Rauscher}, {Rhodes}, {Wang}, {Weinberg}, {Benford}, {Hudson}, {Jeong},
  {Mellier}, {Traub}, {Yamada}, {Capak}, {Colbert}, {Masters}, {Penny},
  {Savransky}, {Stern}, {Zimmerman}, {Barry}, {Bartusek}, {Carpenter}, {Cheng},
  {Content}, {Dekens}, {Demers}, {Grady}, {Jackson}, {Kuan}, {Kruk}, {Melton},
  {Nemati}, {Parvin}, {Poberezhskiy}, {Peddie}, {Ruffa}, {Wallace}, {Whipple},
  {Wollack}, \& {Zhao}}]{RST_2015}
{Spergel}, D., {Gehrels}, N., {Baltay}, C., {et~al.} 2015, arXiv e-prints,
  arXiv:1503.03757, \dodoi{10.48550/arXiv.1503.03757}

\bibitem[{Stefan(1879)}]{Stefan_1879}
Stefan, J. 1879, Sitzungsberichte der Kaiserlichen Akademie der Wissenschaften
  in Wien, 79, 391

\bibitem[{{Stetson}(1996)}]{Stetson96}
{Stetson}, P.~B. 1996, pasp, 108, 851, \dodoi{10.1086/133808}

\bibitem[{{Trentin} {et~al.}(2024){Trentin}, {Ripepi}, {Molinaro}, {Catanzaro},
  {Storm}, {De Somma}, {Marconi}, {Bhardwaj}, {Gatto}, {Testa}, {Musella},
  {Clementini}, \& {Leccia}}]{Trentin_2024_CMetall}
{Trentin}, E., {Ripepi}, V., {Molinaro}, R., {et~al.} 2024, aap, 681, A65,
  \dodoi{10.1051/0004-6361/202347195}

\bibitem[{{Udalski} {et~al.}(1992){Udalski}, {Szymanski}, {Kaluzny}, {Kubiak},
  \& {Mateo}}]{Udalski_1992_OGLE}
{Udalski}, A., {Szymanski}, M., {Kaluzny}, J., {Kubiak}, M., \& {Mateo}, M.
  1992, \actaa, 42, 253

\bibitem[{{Welch} {et~al.}(1987){Welch}, {McLaren}, {Madore}, \&
  {McAlary}}]{Welch_1987_Hband_PLR}
{Welch}, D.~L., {McLaren}, R.~A., {Madore}, B.~F., \& {McAlary}, C.~W. 1987,
  \apj, 321, 162, \dodoi{10.1086/165622}

\bibitem[{{Welch} {et~al.}(1984){Welch}, {Wieland}, {McAlary}, {McGonegal},
  {Madore}, {McLaren}, \& {Neugebauer}}]{Welch84}
{Welch}, D.~L., {Wieland}, F., {McAlary}, C.~W., {et~al.} 1984, apjs, 54, 547,
  \dodoi{10.1086/190943}

\bibitem[{Werner {et~al.}(2004)Werner, Roellig, Low, Rieke, Rieke, Hoffmann,
  Young, Houck, Brandl, Fazio, Hora, Gehrz, Helou, Soifer, Stauffer, Keene,
  Eisenhardt, Gallagher, Gautier, Irace, Lawrence, Simmons, Van~Cleve, Jura,
  Wright, \& Cruikshank}]{Werner_2004_Spitzer}
Werner, M.~W., Roellig, T.~L., Low, F.~J., {et~al.} 2004, The Astrophysical
  Journal Supplement Series, 154, 1, \dodoi{10.1086/422992}

\bibitem[{Wien(1896)}]{Wien_1896}
Wien, W. 1896, Annalen der Physik, 294, 662,
  \dodoi{https://doi.org/10.1002/andp.18962940803}

\bibitem[{{Wisniewski} \& {Johnson}(1968)}]{wisniewski_johnson_1968}
{Wisniewski}, W.~Z., \& {Johnson}, H.~L. 1968, COMMUN. OF THE LUNAR AND
  PLANETARY LAB, 7, 112

\end{thebibliography}

\newpage
\appendix
\counterwithin{figure}{section}
\counterwithin{table}{section}

\section{GLOESS :A fitting method} \label{sec:GLOESS}

GLOESS (Gaussian Localized Scatter Smoothing) \citep{Persson_2004_nearIR} is a statistical technique designed to create a uniform distribution of time or phase, making it suitable for contexts that require time-averaged or mean-phase data. Primarily used in astronomy, this method is particularly effective for smoothing time-series data when dealing with irregularly spaced data points or phases that lack uniform distribution.

This algorithm is a refined version of a smoothing method known as LOESS (Locally Estimated Scatter plot Smoothing), initially introduced by \cite{cleveland_1991_computational}. LOESS applies a local polynomial regression, often a second-order fit, to the data within a defined window. For each window, LOESS assigns a weight to data points based on their distance from the center, giving more influences to points near the midpoint. After fitting the polynomial, an interpolated point is computed at the window’s center to represent the smoothed value for that position. The window then shifts forward by a user-defined increment, and this process repeats across the entire data range, producing a continuous, ``smoothed'' representation that minimizes random variations while preserving overall trends.

In the refined version, instead of utilizing a fixed-size sliding rectangular window, GLOESS employs a scheme that inversely weights the entire dataset. This weighting is based on the statistical uncertainties of the data points and their Gaussian distance from the interpolation point. The size of this ``Gaussian window'' was iteratively adjusted for each light curve to ensure significant high-frequency details were captured without overfitting the data.

\begin{table}[ht]
\caption{Values of the slope of the magnitude–magnitude relation for all possible pairs of filter bands, as illustrated in Figure~\ref{fig:AA_all}.}
\centering
\begin{tabular}{llc}
\hline
\hline
Random-phase & Secondary & Slope  \\
Filter & Filter &\\
\hline 
B & V & $0.893 \pm 0.018$ \\
B & I & $0.815 \pm 0.028$ \\
B & J & $0.744 \pm 0.045$ \\
B & H & $0.721 \pm 0.047$ \\
B & K & $0.708 \pm 0.047$ \\
B & $[$3.6$]$ & $0.694 \pm 0.046$ \\
B & $[$4.5$]$ & $0.715 \pm 0.046$ \\
\hline 
V & B & $1.120 \pm 0.024$ \\
V & I & $0.931 \pm 0.012$ \\
V & J & $0.864 \pm 0.030$ \\
V & H & $0.841 \pm 0.032$ \\
V & K & $0.827 \pm 0.034$ \\
V & $[$3.6$]$ & $0.811 \pm 0.031$ \\
V & $[$4.5$]$ & $0.833 \pm 0.031$ \\
\hline 
I & B & $1.227 \pm 0.043$ \\
I & V & $1.074 \pm 0.015$ \\
I & J & $0.950 \pm 0.017$ \\
I & H & $0.928 \pm 0.020$ \\
I & K & $0.914 \pm 0.021$ \\
I & $[$3.6$]$ & $0.874 \pm 0.025$ \\
I & $[$4.5$]$ & $0.911 \pm 0.023$ \\
\hline 
\end{tabular}
\label{tab:correction_par_1}
\end{table}

\begin{table}[ht]
\caption{Continued from Table \ref{tab:correction_par_1}}
\centering
\begin{tabular}{llc}
\hline
\hline
Random-phase & Secondary & Slope  \\
Filter & Filter & \\
\hline 
J & B & $1.344 \pm 0.078$ \\
J & V & $1.158 \pm 0.043$ \\
J & I & $1.052 \pm 0.029$ \\
J & H & $0.982 \pm 0.005$ \\
J & K & $0.968 \pm 0.006$ \\
J & $[$3.6$]$ & $0.966 \pm 0.010$ \\
J & $[$4.5$]$ & $0.990 \pm 0.009$ \\
\hline 
H & B & $1.387 \pm 0.086$ \\
H & V & $1.189 \pm 0.049$ \\
H & I & $1.077 \pm 0.032$ \\
H & J & $1.018 \pm 0.006$ \\
H & K & $0.987 \pm 0.002$ \\
H & $[$3.6$]$ & $0.984 \pm 0.005$ \\
H & $[$4.5$]$ & $1.007 \pm 0.004$ \\
\hline 
K & B & $1.413 \pm 0.090$ \\
K & V & $1.210 \pm 0.052$ \\
K & I & $1.094 \pm 0.034$ \\
K & J & $1.032 \pm 0.008$ \\
K & H & $1.013 \pm 0.002$ \\
K & $[$3.6$]$ & $0.994 \pm 0.004$ \\
K & $[$4.5$]$ & $1.017 \pm 0.004$ \\
\hline 
$[$3.6$]$ & B & $1.441 \pm 0.089$ \\
$[$3.6$]$ & V & $1.233 \pm 0.052$ \\
$[$3.6$]$ & I & $1.145 \pm 0.036$ \\
$[$3.6$]$ & J & $1.035 \pm 0.016$ \\
$[$3.6$]$ & H & $1.017 \pm 0.012$ \\
$[$3.6$]$ & K & $1.006 \pm 0.011$ \\
$[$3.6$]$ & $[$4.5$]$ & $1.024 \pm 0.003$ \\
\hline 
$[$4.5$]$ & B & $1.398 \pm 0.084$ \\
$[$4.5$]$ & V & $1.200 \pm 0.050$ \\
$[$4.5$]$ & I & $1.098 \pm 0.033$ \\
$[$4.5$]$ & J & $1.010 \pm 0.015$ \\
$[$4.5$]$ & H & $0.993 \pm 0.012$ \\
$[$4.5$]$ & K & $0.983 \pm 0.011$ \\
$[$4.5$]$ & $[$3.6$]$ & $0.976 \pm 0.003$ \\
\hline 
\end{tabular}
\label{tab:correction_par_2}
\end{table}
\clearpage
\begin{figure}
    \vspace{2cm}
    \centering
    \includegraphics[width=1\textwidth]{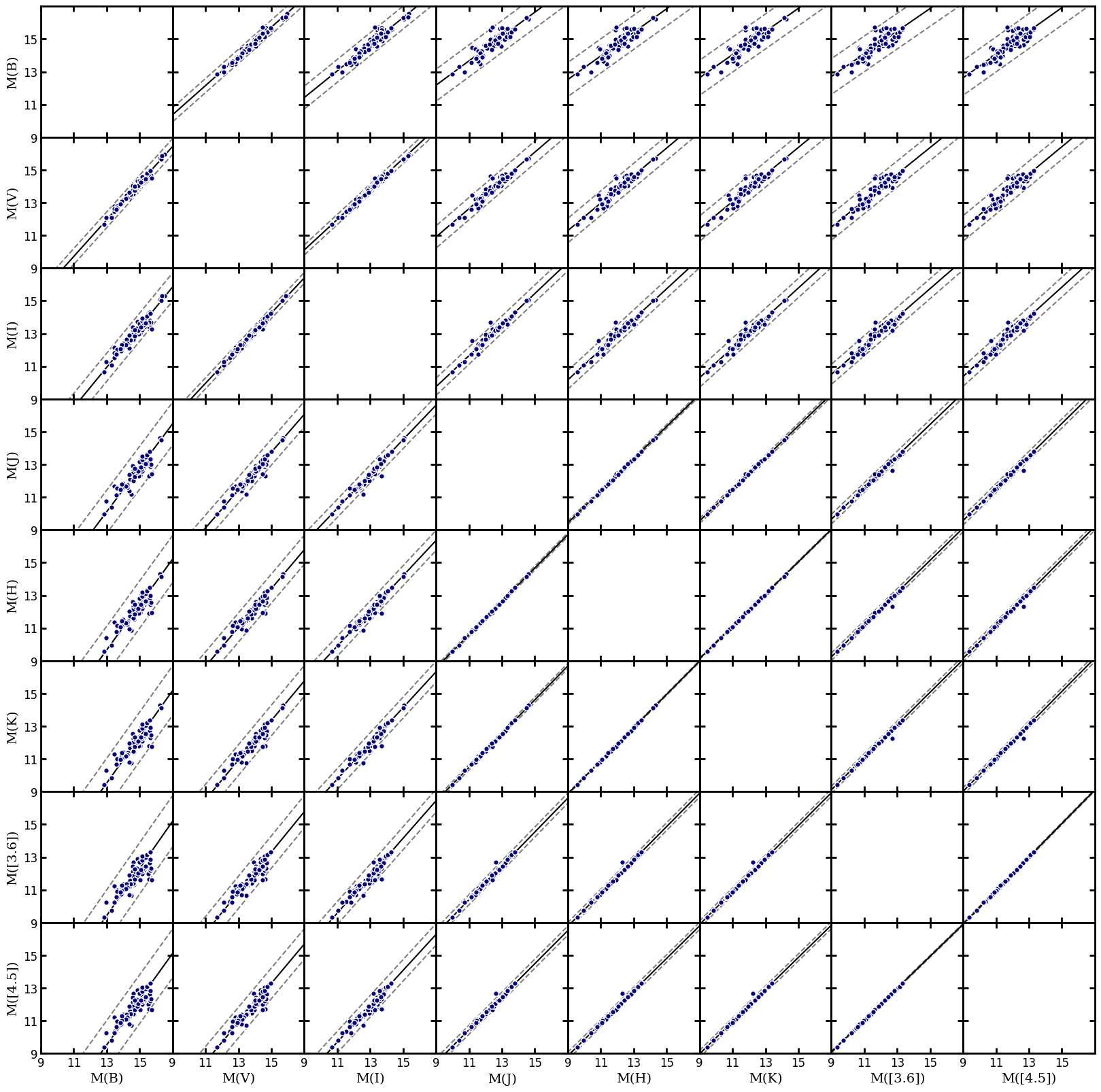}
    \caption{Pair plots of magnitude-magnitude in different bands show that consecutive filter bands exhibit correlations. The solid line represents the fitted data, and the dashed line indicates a 1$\sigma$ dispersion. Notably, dispersion tends to be lower when bands are closer in wavelength. The slopes of all the fitted lines are pretty similar, while the intercepts show significant variations.}
    \label{fig:AA_all}
\end{figure}
\begin{figure}
    \vspace{2cm}
    \centering
    \includegraphics[width=1\textwidth]{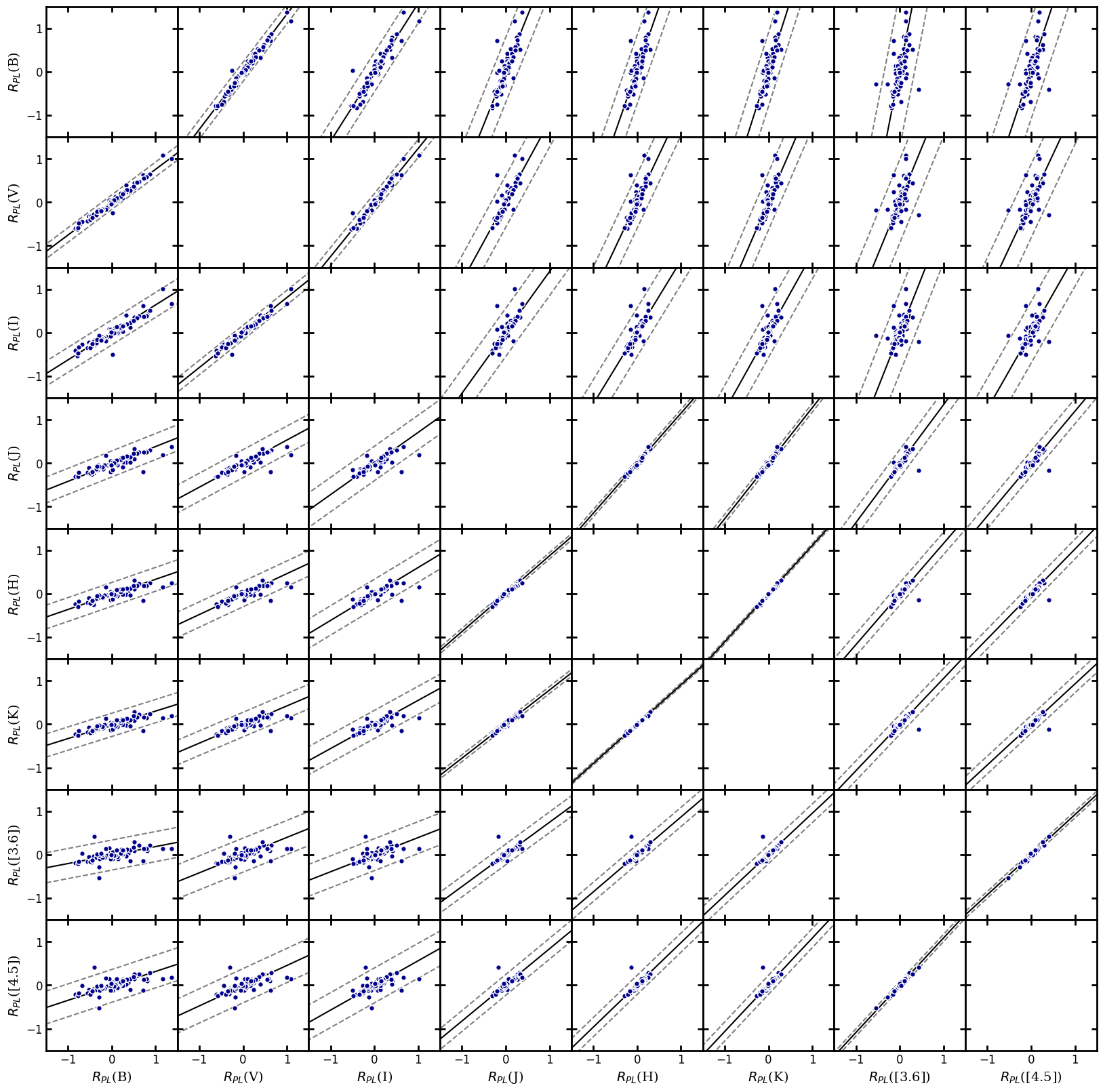}
    \caption{Pair plots illustrate the correlation between residuals from two different bands, both derived from the PL-fitted line. A notable correlation is evident among the residuals, particularly for consecutive filter bands. This study primarily focuses on the $J$ and $B$ bands, which, despite their distinct wavelengths, demonstrate a strong correlation in the (1,4) subplot as indicated by the matrix indices.}
    \label{fig:Pairplot_RR}
\end{figure}
\begin{figure}
    \vspace{2cm}
    \centering
    \includegraphics[width=1\textwidth]{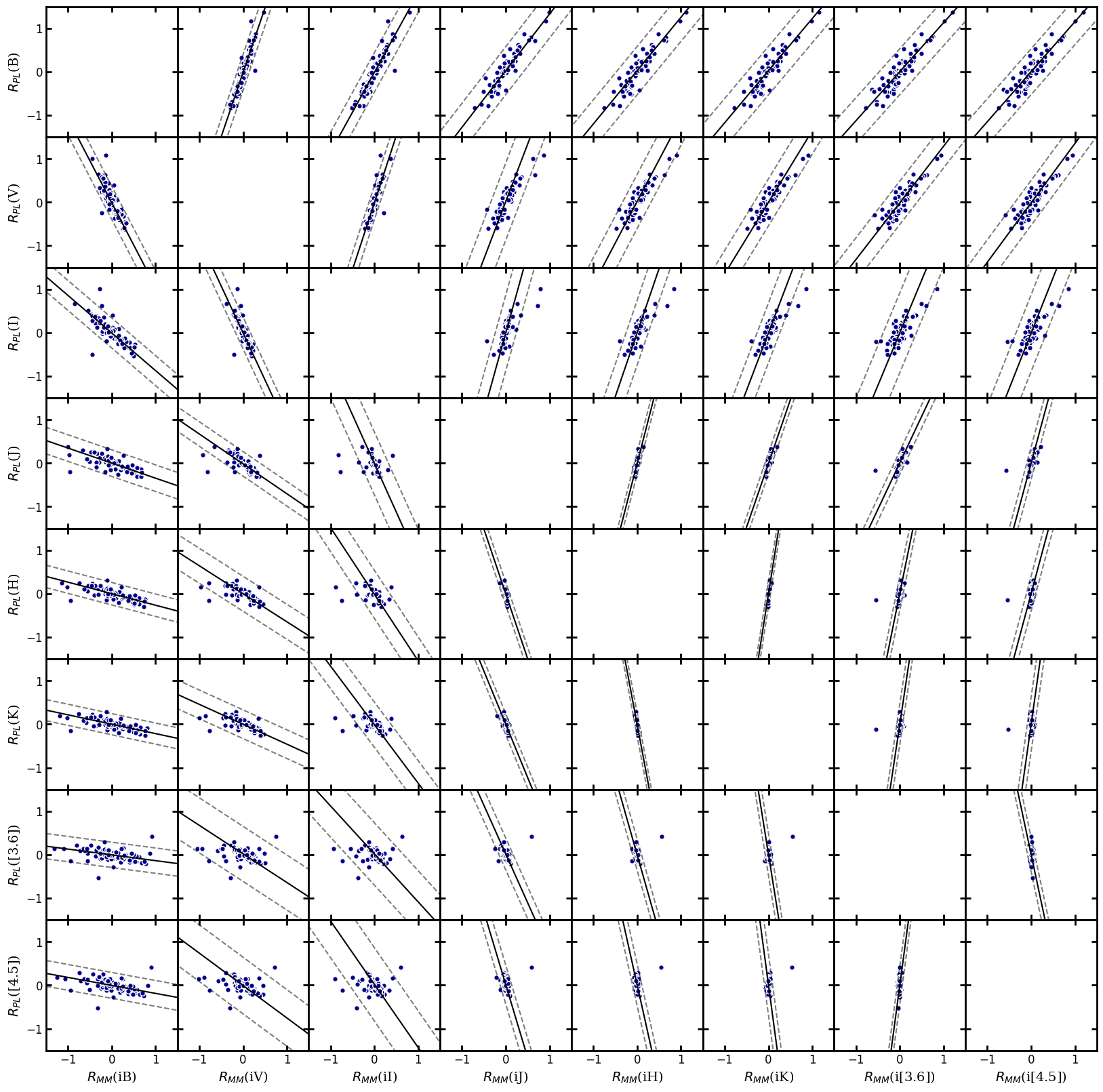}
    \caption{Pair plots present the residuals from the PL-fitted line on the y-axis and the residuals from the magnitude-magnitude fitted line on the x-axis across all bands. It reveals significant correlations between these distinct residuals for each pair of bands. As shown in Figure~\ref{fig:Pairplot_RR}, the primary focus is on the $J$ and $B$ bands, which demonstrate a notable correlation in the (1,4) subplot based on matrix indices. The region corresponding to the combination of infrared (IR) filters is shown in more detail in Figure~\ref{fig:Pairplot_AA_IR}}
    \label{fig:Pairplot_AA}
\end{figure}
\begin{figure}
    \vspace{2cm}
    \centering
    \includegraphics[width=0.8\textwidth]{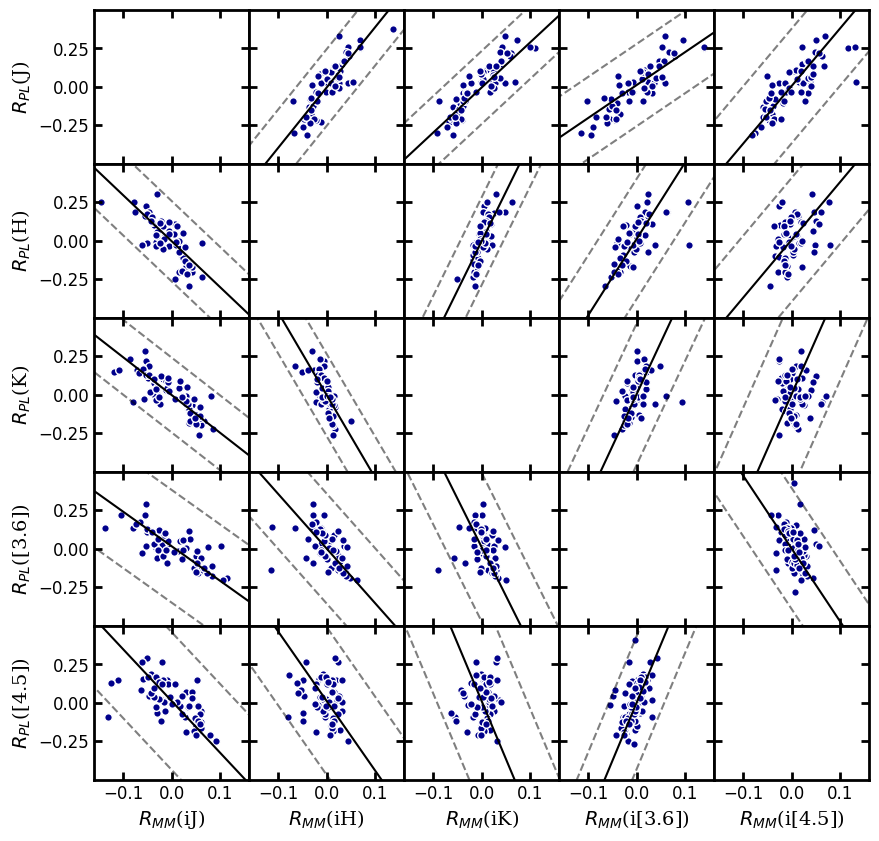}
    \caption{ Enlarged view of the region in Figure~\ref{fig:Pairplot_AA} corresponding to combinations of infrared (IR) bands, illustrating the residual correlations between PL- and magnitude–magnitude-fitted relations. Axis limits have been rescaled relative to Figure~\ref{fig:Pairplot_AA} to enhance the visibility of structural features in the distributions.}
    \label{fig:Pairplot_AA_IR}
\end{figure}
\begin{figure}
   \vspace{1cm}
    \centering
    \includegraphics[width=0.85\textwidth]{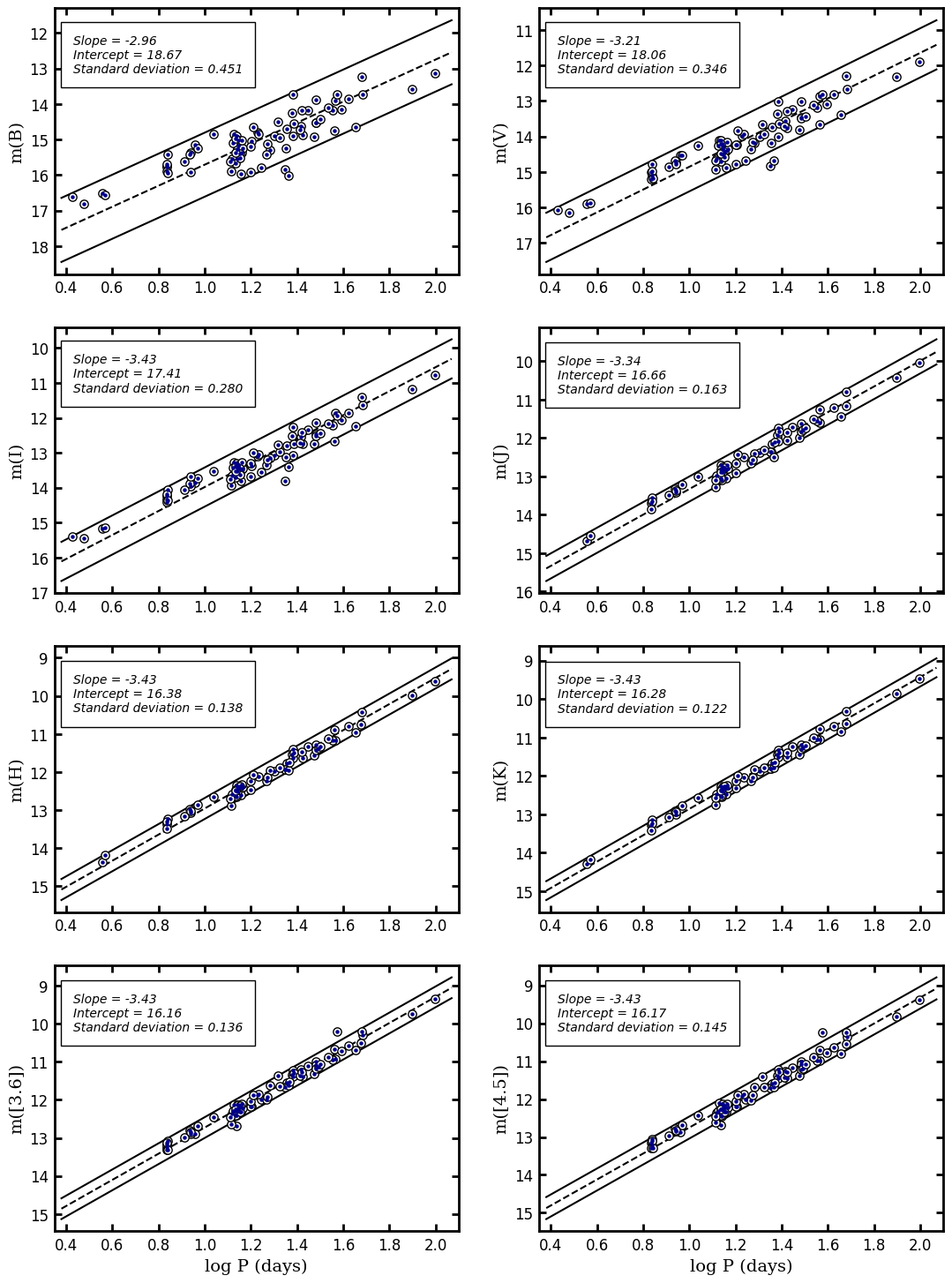}
    \caption{Similar to Figure~\ref{fig:PLR_filters_ALLinONE}, but presented as separate figures. Each figure includes the slope and intercept of the fitted line, along with dispersion information represented by 2$\sigma$ solid black lines.}
    \label{fig:PLR_filters}
\end{figure}

\end{document}